\documentclass{emulateapj}

\usepackage{graphicx}
\usepackage{graphics}

\def\gsim{\;\lower4pt\hbox{${\buildrel\displaystyle >\over\sim}$}\,}
\def\lsim{\;\lower4pt\hbox{${\buildrel\displaystyle <\over\sim}$}\,}

\def\FLASH{{\sc flash}}

\slugcomment{}

\shorttitle{Supernova 1987A: a Template to Link Supernovae to their
     Remnants}
\shortauthors{S. Orlando et al.} 

\begin{document}

\newcommand\rs[1]{_\mathrm{#1}}

\title{Supernova 1987A: a Template to Link Supernovae to their Remnants}

\author{S. Orlando\altaffilmark{1}, M. Miceli\altaffilmark{2,1},
        M.L. Pumo\altaffilmark{1,3,4},
        F. Bocchino\altaffilmark{1}}
\email{orlando@astropa.inaf.it}

\altaffiltext{1}{INAF - Osservatorio Astronomico di Palermo ``G.S.
              Vaiana'', Piazza del Parlamento 1, 90134 Palermo, Italy}
\altaffiltext{2}{Dip. di Fisica e Chimica, Univ. di Palermo, Piazza del Parlamento 1, 90134 Palermo,
              Italy}
\altaffiltext{3}{INAF -- Osservatorio Astronomico di Padova, Vicolo
              dell'Osservatorio 5, 35122 Padova, Italy}
\altaffiltext{4}{INAF -- Osservatorio Astrofisico di Catania, Via S. Sofia
              78, 95123 Catania, Italy}

\begin{abstract}
The emission of supernova remnants reflects the properties of both
the progenitor supernovae and the surrounding environment. The complex
morphology of the remnants, however, hampers the disentanglement of the
two contributions. Here we aim at identifying the imprint of SN\,1987A on
the X-ray emission of its remnant and at constraining the structure of
the environment surrounding the supernova. We performed high-resolution
hydrodynamic simulations describing SN\,1987A soon after the core-collapse
and the following three-dimensional expansion of its remnant between
days 1 and 15000 after the supernova. We demonstrated that the physical
model reproducing the main observables of SN\,1987A during the first 250
days of evolution reproduces also the X-ray emission of the subsequent
expanding remnant, thus bridging the gap between supernovae and supernova
remnants. By comparing model results with observations, we constrained
the explosion energy in the range $1.2-1.4\times 10^{51}$~erg and the
envelope mass in the range $15-17 M_{\odot}$. We found that the shape
of X-ray lightcurves and spectra at early epochs ($<15$ years) reflects
the structure of outer ejecta: our model reproduces the observations
if the outermost ejecta have a post-explosion radial profile of density
approximated by a power law with index $\alpha = -8$. At later epochs,
the shapes of X-ray lightcurves and spectra reflect the density structure
of the nebula around SN\,1987A.  This enabled us to ascertain the origin
of the multi-thermal X-ray emission, to disentangle the imprint of
the supernova on the remnant emission from the effects of the remnant
interaction with the environment, and to constrain the pre-supernova
structure of the nebula.
\end{abstract}


\section{Introduction}
\label{sec1}

Supernova remnants (SNRs), the leftovers of supernova (SN) explosions,
are diffuse extended sources with a quite complex morphology.
General consensus is that such morphology reflects, on one hand,
the physical and chemical properties of the progenitor SN and, on
the other hand, the early interaction of the SN blast wave with the
inhomogeneous circumstellar medium (CSM). Thus investigating the
intimate link that exists between the morphological properties of
a SNR and the complex phases in the SN explosion may help: 1) to
trace back the structure and chemical composition of SN ejecta, and
the dynamics and energetics of the SN explosion; 2) to probe the
structure and geometry of the CSM, thereby mapping the final stages
of the stars evolution.  However, the very different time and space
scales of SNe and SNRs make difficult to study in detail their
connection (e.g. \citealt{2008ApJ...680.1149B, 2012ApJ...755..160E,
2014ApJ...785L..27Y, 2015ApJ...803..101P}).

Because of its youth and proximity, SN\,1987A in the Large Magellanic
Cloud offers a unique opportunity to bridge the gap between the
progenitor SN and its remnant. SN\,1987A was an hydrogen-rich
core-collapse (CC) SN that was observed in outburst on February 23,
1987 (\citealt{1987A&A...177L...1W}). It occurred approximately
51.4 kpc from Earth (\citealt{1999IAUS..190..549P}). About
80 days after the explosion, it reached a peak apparent visual
magnitude of $\approx 3$ (naked eye visible; e.g.
\citealt{1988MNRAS.231P..75C, 1988AJ.....95...63H}). Its evolution
has been accurately monitored in different wavelength bands since
the outburst, from infrared (e.g. with {\it Spitzer},
\citealt{2010ApJ...722..425D}), to optical (e.g. with {\it Hubble
Space Telescope}, \citealt{2000ApJ...537L.123L, 2011Natur.474..484L}),
to X-ray bands (e.g. with {\it Rosat}, \citealt{2006A&A...460..811H},
{\it Chandra}, \citealt{2013ApJ...764...11H}, and {\it XMM-Newton},
\citealt{2006A&A...460..811H}). This has provided a wealth of
high-quality data with unprecedented completeness, making of
SN\,1987A an ideal template to study the SN-SNR connection.

The study, however, is complicated by the interaction of the blast wave
with the surrounding inhomogeneous medium. Optical images soon after the
outburst revealed an enigmatic triple-ring nebula around the SN
(\citealt{1989ApJ...347L..61C}). The nebula consists mainly of a dense
central equatorial ring and two outer rings displaced by about 0.4~pc
above and below the central ring and lying in planes almost parallel to
the equatorial one. It has been suggested that the nebula might be the
result either of the merging of two massive stars occurred some 20,000
years before the explosion (\citealt{2007Sci...315.1103M}) or of mass loss
from a fast-rotating star (\citealt{2008A&A...488L..37C}).

The interaction of the SN with the nebula is best observed
in the radio and X-ray bands. It started about 3 years after the
explosion when both radio and X-ray fluxes began to increase with
time (\citealt{1996A&A...312L...9H,1997ApJ...479..845G}). This
was interpreted as due to outer ejecta lighting up the dense wind
emitted during a previous red supergiant (RSG) phase and subsequently
swept-up by the fast wind during a phase of blue supergiant (BSG)
(\citealt{1995ApJ...452L..45C}). After about 16 years, suddenly the
soft X-ray lightcurve has steepened still further, contrary to the
hard X-ray and radio lightcurves (\citealt{2005ApJ...634L..73P}). This
was interpreted as due to the blast wave sweeping up the dense central
equatorial ring (\citealt{2007AIPC..937....3M}). Since then the soft
X-ray flux continued to grow, indicating that currently the shock
is still traveling through the dense part of the equatorial ring
(\citealt{2013ApJ...764...11H}).

X-ray observations more than others encode information about the
physical properties of both the nebula and the stellar ejecta.
Decoding these observations, therefore, might open the possibility
to reconstruct the nebula and ejecta structures soon after the SN
explosion. This requires accurate and detailed numerical models
able to follow the ejecta evolution from the SN explosion to the
SNR phase and to reproduce altogheter the emission properties of
both the progenitor SN and its remnant.

Current models, however, were usually aimed at describing either
the SN evolution (e.g. \citealt{2011ApJ...741...41P, 2014ApJ...783..125H})
or the expansion of the remnant (e.g. \citealt{2009ApJ...692.1190Z,
2012ApJ...752..103D, 2014ApJ...794..174P}). The former models
describe the early SN evolution and ignore its subsequent interaction
with the nebula; the latter assume an initial parametrized ejecta
profile that is not proven to reproduce the main observables of the
progenitor SN, leaving out an accurate description of the ejecta
distribution of mass and energy soon after the SN explosion. These
limits make difficult to disentangle the effects of the
initial conditions (i.e. the SN event) from those of the boundary
conditions (i.e. the interaction with the environment). A first
attempt to connect some properties of CC-SN (e.g. the composition
and a parameterization of the circumstellar environment) to some
observable bulk properties of SNR has been discussed recently
(\citealt{2015ApJ...803..101P}), even though by adopting a 1D
approach.

The structure of the pre-SN nebula of SN\,1987A is inherently
three-dimensional (3D), as is visible in the images from the Hubble
Space Telescope (HST) (e.g. \citealt{2011Natur.474..484L}). A proper
description of the interaction of the ejecta with the nebula therefore
requires 3D calculations. To date only a 3D model has been proposed
(\citealt{2014ApJ...794..174P}) to explore the origin of the asymmetric
radio morphology observed in SN\,1987A. However this model focusses
on the analysis of the radio emission and its initial condition is a
parametrized function describing the ejecta distribution about two years
after the explosion. As a consequence their model cannot describe the
evolution of the SN and their adopted distribution of ejecta is not
proven to be consistent with the observables of the SN.

Here we present a hydrodynamic model describing the evolution of
SN\,1987A from the breakout of the shock wave at the stellar surface
until the expansion of its remnant through the nebula surrounding the
SN. We ran several high-resolution simulations to reproduce altogether
the main observables of the SN (i.e. bolometric lightcurve, evolution
of line velocities, and continuum temperature at the photosphere)
and the properties of X-ray emission of the remnant (lightcurves,
spectra, and morphology). Our simulations cover the first 40 years of
evolution to make predictions on the future evolution of the remnant
in view of the spatially-resolved high-resolution spectroscopy
capability of the forthcoming X-ray observatory {\it Athena}
(\citealt{2013arXiv1306.2307N}). The paper is organized as follows. In
Section~\ref{sec2} we describe the hydrodynamic model and the numerical
setup, in Section~\ref{sec3} we describe the results and, finally,
we draw our conclusions in Section~\ref{sec4}.

\section{Problem description and numerical setup}
\label{sec2}

We first modeled the SN by performing one-dimensional (1D) simulations
of the ``early'' post-explosion evolution of a CC-SN during
the first 250 days (see Section~\ref{modelSN}). Then the output of these
simulations was mapped into 3D and provided the initial condition for
the ejecta structure of a 3D model describing the further evolution of
the SN and the subsequent development of the SNR interacting with the
CSM (see Section~\ref{modelSNR}).

\subsection{Modeling the post-explosion evolution of the supernova}
\label{modelSN}

The post-explosion evolution of the SN was modeled by using a 1D
Lagrangian code, specifically tailored to simulate the main observables in
CC-SNe (namely the bolometric lightcurve and the time evolution
of the photospheric velocity and temperature; \citealt{2011ApJ...741...41P})
and widely used to model observed events (\citealt{2014MNRAS.439.2873S,
2014MNRAS.438..368T, 2014ApJ...787..139D}) including the SN 1987A-like
ones (\citealt{2012A&A...537A.141P}). 

The code solves the equations of relativistic radiation hydrodynamics
in spherical symmetry for a self-gravitating matter fluid which
interacts with radiation. Its distinctive features are: 1) a fully
general relativistic treatment; 2) an accurate treatment of radiative
transfer at all regimes (from the one in which the ejecta are
optically thick up to when they are optically thin); 3) the coupling
of the radiation moment equations with the equations of relativistic
hydrodynamics during all the post-explosive phases, adopting a fully
implicit Lagrangian finite difference scheme; 4) the computation
of heating effects due to decays of radioactive isotopes synthesized
in the SN explosion; and 5) the computation of the gravitational
effects of the central compact object on the evolution of the
ejecta.

Thanks to these characteristics, the code is able to compute the
evolution of the stellar ejecta and emitted luminosity from the
breakout of the shock wave at the stellar surface up to the so-called
nebular stage (i.e. when the envelope has recombined and the energy
budget is dominated by the radioactive decays of the heavy elements
synthesized in the explosion, see also Section~\ref{evolSN}). At the same
time, the code is able to accurately determine the fallback of
material on the central object and, as a consequence, the amount
of $^{56}$Ni present in the ejected envelope at late times.

The simulations start from the breakout of the shock wave at the stellar
surface, using initial conditions that mimic the physical properties of
the ejected material after shock passage following the core-collapse
(\citealt{2011ApJ...741...41P}).  As a consequence, the models depend
on some basic parameters that characterize the main radiative, thermal,
and dynamical properties of such material. These parameters are: the
progenitor radius $R_0$, the total ejecta energy  $E_{\rm SN}$, the
envelope mass at shock breakout $M_{\rm env}$, and the total amount of
$^{56}$Ni initially present in the ejected envelope $M_{\rm Ni}$.  In our
simulations $M_{\rm env}$ coincides with the total mass surrounding
the central compact object at the start of simulations.  So, during
the post-explosive evolution, $M_{\rm env}$ is equal to the ejected
mass $M_{\rm ej}$ plus the mass fallen back to the central compact
object. In our models of SN\,1987A, the latter mass is of the order of
a few hundredths of a solar mass and, as a consequence, $M_{\rm env}$
essentially corresponds to $M_{\rm ej}$.

We performed several simulations of the SN evolution, exploring
the parameter space defined by $M_{\rm env}$ and $E_{\rm SN}$.
The exploration has been restricted to the ranges of values commonly
discussed in the literature for SN\,1987A (\citealt{1987ApJ...319..136A,
1988ApJ...330..218W, 2005A&A...441..271U, 2011ApJ...741...41P,
2014ApJ...783..125H}): we considered models with $M_{\rm env}$ ranging
between 8 and 19~$M_{\odot}$ (including the bulk of ejecta and the
high-velocity shell), and $E_{\rm SN}$ ranging between 1 and $2\times
10^{51}$~erg. Also, in our model, the outermost high-velocity shell ($4000
< v < 20000$~km~s$^{-1}$; \citealt{2000ApJ...537L.123L}) can be described
by a power law with index $\alpha$ (\citealt{1990ApJ...360..242S,
2000ApJ...532.1132B}). Here we explored models with $\alpha$ ranging
between $-8$ and $-9$, according to the values suggested in the
literature (e.g. \citealt{1988ApJ...330..218W}).  We fixed all the
other parameters of the model, namely the initial radius $R_0 = 3\times
10^{12}$~cm (which is a reliable value for the progenitor of SN\,1987A;
\citealt{arnett96, 2004ApJ...617.1233Y, 2011ApJ...741...41P}) and the
initial amount of $^{56}$Ni $M_{Ni} = 0.07 M_{\odot}$ (corresponding
to the amount of Ni synthesized during the explosion of SN 1987A;
\citealt{1989ARA&A..27..629A}).

Note that our 1D SN model cannot simulate possible asymmetries
developed during the explosion as suggested by a significant body
of observational and simulation evidence. It goes without saying
that the most desirable way to model a CC-SN would include a
multi-dimensional hydrodynamical calculation, following the
core-collapse, the bounce at nuclear densities, the leakage of
neutrinos from the proto-neutron star, the neutrino heating, and
the delayed explosion with the shock propagating through the envelope
to be ejected. However, current multi-dimensional models can follow
the evolution of the exploding star only for a very short
time\footnote{However note that a new approach to follow
the post-explosion evolution of the SN in multi-dimensions for a
longer time has been proposed recently by \cite{2009ApJ...693.1780J}.}
(of the order of few minutes), preventing any possibility to use
their output for post-explosion calculations that follow in detail
the evolution of the CC-SN ejecta. Conversely, our 1D model allows
us to follow the post-explosion evolution of the SN for days, to
describe in detail the fallback by using a full-relativistic approach
and, therefore, to estimate accurately the total mass ejected in
the SN explosion.  Also our lagrangian code is not flux-limited (as
the multi-dimensional codes are), allowing us to simulate accurately
the evolution of the ejected material during both the initial phase
of the shock breakout and the following nebular phase in which the
ejected SN envelope has recombined.


\subsection{Modeling the supernova remnant evolution}
\label{modelSNR}

Optical images clearly show that the structure of the
nebula as well as the morphology and evolution of the
ejecta are inherently 3D (\citealt{2010A&A...517A..51K,
2011Natur.474..484L,2013ApJ...768...89L}). Therefore, once the early phase
of the SN explosion has been modeled in 1D (see Section~\ref{modelSN}),
the output of the SN simulation was mapped into 3D. Then the subsequent
evolution and transition from the SN phase to the SNR phase were modeled
by numerically solving the time-dependent fluid equations of mass,
momentum, and energy conservation; the equations in a 3D Cartesian
coordinate system $(x,y,z)$ take into account the radiative losses from
an optically thin plasma:

\begin{equation}
\begin{array}{l}\displaystyle
\frac{\partial \rho}{\partial t} + \nabla \cdot \rho \mbox{\bf u} = 0~,
\\ \\ \displaystyle
\frac{\partial \rho \mbox{\bf u}}{\partial t} + \nabla \cdot \rho
\mbox{\bf uu} + \nabla P = 0~,
\\ \\ \displaystyle
\frac{\partial \rho E}{\partial t} +\nabla\cdot (\rho E+P)\mbox{\bf u}
= -n_{\rm e} n_{\rm H} \Lambda(T)~,
\end{array}
\label{mod_eq}
\end{equation}

\bigskip
\noindent
where $E = \epsilon + |\mbox{\bf u}|^2/2$ is the total gas energy
(internal energy, $\epsilon$, and kinetic energy), $t$ is the time,
$\rho = \mu m_H n_{\rm H}$ is the mass density, $\mu = 1.3$ is the
mean atomic mass (assuming cosmic abundances), $m_H$ is the mass
of the hydrogen atom, $n_{\rm H}$ is the hydrogen number density,
$n_{\rm e}$ is the electron number density, {\bf u} is the gas
velocity, $T$ is the temperature, and $\Lambda(T)$ represents the
radiative losses per unit emission measure. We used the ideal gas
law, $P=(\gamma-1) \rho \epsilon$, where $\gamma=5/3$ is the adiabatic
index.

The calculations were performed using the {\FLASH} code
(\citealt{for00}). The hydrodynamic equations were solved using the
{\FLASH} implementation of the Piecewise Parabolic Method (PPM) algorithm
(\citealt{cole84}).  For the present application, the code has been
extended by additional computational modules to handle the radiative
losses $\Lambda$ (through a table lookup/interpolation method) and to
calculate the deviations from electron-proton temperature-equilibration
and the deviations from equilibrium of ionization of the most abundant
ions.  The output of the latter modules was used in the synthesis of
X-ray emission (see Section~\ref{synthesisSNR}).

The 3D simulations start once the majority of the energy released
in the explosion was kinetic ($\sim 26$ hours after the explosion).
Note that, at variance with our 1D SN model, the 3D
simulations do not include a heating term due to decays of radioactive
isotopes synthesized in the SN explosion, such as $^{56}$Co or
$^{44}$Ti. On the other hand, in Appendix~\ref{SNSNRcomparison},
we show that our simplification does not affect significantly the
radial profiles of density and velocity of the ejecta at later
times. This evidence supports our assumption to neglect the effect
of radioactive heating during the interaction of the remnant with
the H\,II region and with the dense equatorial ring.

We followed the interaction of the blast wave and ejecta with the
circumstellar nebula during the first 40 years of evolution (namely
until the presumed launch date of the {\it Athena} X-ray
observatory; \citealt{2013arXiv1306.2307N}). The initial remnant radius
was $\approx 20$~AU ($\approx 10^{-4}$~pc). As suggested by previous
studies (\citealt{2012ApJ...749..156O}), we assumed that the initial density
structure of the ejecta was clumpy. To this end, after the
1D radial density distribution of ejecta was mapped into 3D, we modeled
the ejecta clumps as per-cell random density perturbations derived
from a power-law probability distribution (\citealt{2012ApJ...749..156O}).
The latter distribution, with index $\xi = -1$, is characterized
by a parameter $\nu_{\rm max}$ representing the maximum density
perturbation allowed in the simulation. For the purposes of this
paper, we assumed that the ejecta clumps have initial size $6\times
10^{12}$~cm (0.4~AU, corresponding to 2\% of the initial remnant
radius) and a maximum density perturbation $\nu_{\rm max} = 5$. The
size and density contrast of the modeled ejecta clumps are in
agreement with those suggested by spectropolarimetric studies of
SNe (\citealt{2003ApJ...591.1110W, 2004ApJ...604L..53W, 2010ApJ...720.1500H}).

\begin{table*}
\caption{Adopted parameters of the CSM for the hydrodynamic models of the SN\,1987A
event.}
\label{tab1}
\begin{center}
\begin{tabular}{lclcc}
\hline
\hline
CSM component & Parameters     & Units & Range of values explored &  Best-fit values  \\
\hline
BSG wind: & $\dot{M}_{\rm w}$  & ($M_\odot$~year$^{-1}$) &  $10^{-7}$ & $10^{-7}$     \\
  & $v_{\rm w}$    &  (km~s$^{-1}$)   &  500          & 500    \\
  & $r_{\rm w}$    &  (pc)            &  0.05         & 0.05   \\
\hline
H\,II region: & $n_{\rm HII}$  &  ($10^2$ cm$^{-3}$) &  $[0.8-3]$   & 0.9     \\
 & $r_{\rm HII}$  &  (pc)            &  $[0.08-0.2]$   & 0.08    \\
\hline
Equatorial ring: & $n_{\rm rg}$   &  ($10^3$ cm$^{-3}$) &  $[1-2]$     & 1  \\
 & $r_{\rm rg}$   &  (pc)            &  0.18         & 0.18    \\
 & $w_{\rm rg}$   &  ($10^{17}$ cm)  &  $[0.7-2]$      & $1.7$   \\
 & $h_{\rm rg}$   &  ($10^{16}$ cm)  &  $3.5$        & $3.5$   \\
\hline
Clumps: & $<n_{\rm cl}>$ &  ($10^4$ cm$^{-3}$) &  $[1-3]$     & $2.5\pm 0.3$      \\
 & $<r_{\rm cl}>$ &  (pc)            &  $[0.14-0.17]$  & $0.155\pm 0.015$     \\
 & $w_{\rm cl}$   &  ($10^{16}$ cm)  &  $[1-3]$        & $1.7$           \\
 & $N_{\rm cl}$   &                  &  $[40-70]$      &   50         \\
\hline
\end{tabular}
\end{center}
\end{table*}

Initially the blast wave from the SN explosion propagates through the
wind of the progenitor BSG. We adopted wind values
discussed in the literature (\citealt{2007Sci...315.1103M}); in particular,
we assumed a spherically symmetric wind with a mass-loss rate of
$\dot{M}_{\rm w} = 10^{-7} M_\odot$~year$^{-1}$ and wind velocity
$v_{\rm w} = 500$~km~s$^{-1}$; the wind gas density is proportional
to $r^{-2}$ (where $r$ is the radial distance from SN\,1987A) and the
termination shock of the wind is located approximately at $r_{\rm w} =
0.05$~pc. Table~\ref{tab1} summarizes the parameters adopted.

\begin{figure}[!t]
  \centering
  \includegraphics[width=8.5cm]{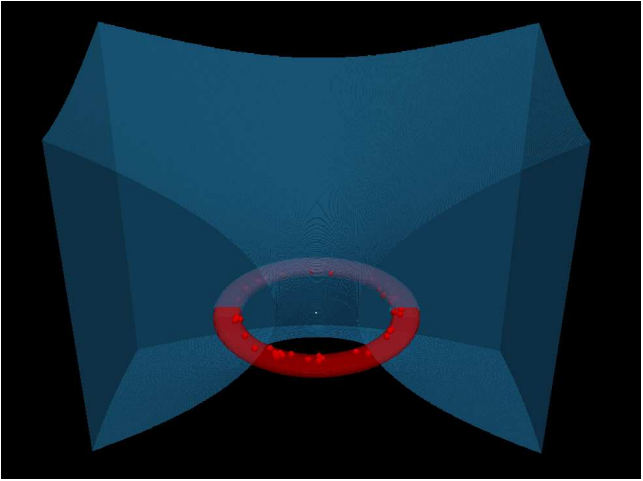}
  \caption{A rendition in log scale of the circumstellar nebula around
  SN\,1987A model initial conditions. The ring consisting of a uniform
  smooth component and high-density spherical clumps is shown in red; the H\,II
  region around the ring is marked by the blue clipped component.
  The white dot at the center of the coordinate system shows the
  position of the SN explosion.}
\label{ini_cond}
\end{figure}

The circumstellar nebula probably originates from the interaction of a
slow wind from an early RSG phase with a faster wind from a subsequent
BSG phase (\citealt{1978ApJ...219L.125K}).  Thus the nebula was modeled
assuming that it is composed by a dense inhomogeneous equatorial ring
surrounded by an ionized H\,II region (\citealt{1995ApJ...452L..45C,
2005ApJS..159...60S}).  Figure~\ref{ini_cond} shows an example of
initial configuration of the nebula. The ring (marked red in the
figure) consists of a uniform smooth component and high-density
spherical clumps mostly located in the inner portion of the ring.
The smooth component has an elliptical cross section with the major
axis $w_{\rm rg}$ lying on the equatorial plane. The clumps mimic the
protrusions emanating from the equatorial ring and probably formed
by hydrodynamic instabilities caused by the interaction of a BSG and
a RSG wind (\citealt{2005ApJS..159...60S}). The clumps have all the
same diameter $w_{\rm cl}$ but their plasma density and radial distance
from SN\,1987A are randomly distributed around the values $<n_{\rm cl}>$
and $<r_{\rm cl}>$ respectively.  The 3D shape and geometry of the H\,II
region were assumed to be analogous to those suggested from the analysis
of HST data of the ring nebula around the BSG SBW1, a candidate twin of
the SN\,1987A progenitor (\citealt{2013MNRAS.429.1324S}).

We explored the parameter space defined by: a) the plasma
density $n_{\rm HII}$ and inner edge $r_{\rm HII}$ of the H\,II
region; b) the density $n_{\rm rg}$, radial distance from SN\,1987A
$r_{\rm rg}$, radial extension $w_{\rm rg}$, and height $h_{\rm rg}$
of the uniform component of the ring; c) the average density $<n_{\rm
cl}>$, average radial distance from SN\,1987A $<r_{\rm cl}>$, diameter
$w_{\rm cl}$, and number $N_{\rm cl}$ of high-density spherical
clumps of the ring. In order to limit as much as possible our
exploration (that is very computer demanding in the case of a 3D
model), we adopted as fiducial values those reported in the literature
and found by comparing the results of a 1D hydrodynamic model of
SN\,1987A with observations (\citealt{2012ApJ...752..103D}). Then we explored
the parameter space around these values, adopting an iterative process
of trial and error to converge on model parameters that reproduce the
X-ray lightcurve and spectra of SN\,1987A. Table~\ref{tab1} summarizes
the ranges of values explored and the parameters of the model best-reproducing
the data.

We traced the evolution of the different plasma components (namely the
ejecta, the H\,II region, and the ring material) by considering passive
tracers associated with each of them. Each material is initialized with
$C_{\rm i} = 1$, while $C_{\rm i} = 0$ elsewhere, where the index ``i''
refers to the ejecta (ej), the H\,II region (HII), and the ring material
(rg). During the remnant evolution, the different materials mix together,
leading to regions with $0 < C_{\rm i} < 1$. At any time $t$ the density
of a specific material in a fluid cell is given by $\rho_{\rm i} =
\rho C_{\rm i}$.

We assumed the SN explosion at the origin of the 3D Cartesian
coordinate system $(x_0, y_0, z_0) = (0, 0, 0)$. The system is
oriented in such a way that the dense equatorial ring lies on the
$(x,y)$ plane. The computational domain extends 1~pc in the $x$,
$y$, and $z$ directions. A major challenge in modeling the explosion
and subsequent evolution of SN\,1987A was the very small scale of
the system in the immediate aftermath of the SN explosion (the
initial remnant radius is $\approx 10^{-4}$~pc) with the size of
the rapidly expanding blast wave. To capture this range of scales,
we exploited the adaptive mesh refinement capabilities of the {\FLASH}
code, employing 18 nested levels of refinement, with resolution
increasing twice at each refinement level. The refinement/derefinement
criterion (\citealt{loehner}) implemented in {\FLASH} follows the changes
in mass density, temperature, tracer of ejecta, and tracer of the ring.

The calculations were performed using also an automatic mesh
derefinement scheme in the whole spatial domain that kept the
computational cost approximately constant as the blast expanded
(\citealt{2012ApJ...749..156O}): the maximum number of refinement levels
used in the calculation gradually decreased from 18 (initially) to 9
(at the final time) following the expansion of the blast and ensuring a
number of grid zones per radius of the remnant $> 100$ during the whole
evolution. At the beginning (at the end) of the simulation, this grid
configuration yielded an effective resolution of $\approx 10^{-6}$~pc
($\approx 5\times 10^{-4}$~pc) at the finest level, corresponding to
$\approx 100$ zones per initial radius of the remnant ($> 600$ zones
per final radius of the remnant). The effective mesh size varied from
$(10^6)^3$ initially to $2048^3$ at the final time. Note that the maximum
spatial resolution achieved by \cite{2014ApJ...794..174P} in their 3D
model was $\approx 4\times 10^{-3}$~pc (corresponding to an effective
mesh size of $256^3$) during the whole evolution.

The high-spatial resolution achieved in our simulations required about
seven millions of CPU hours on the MareNostrum III cluster hosted at the
Barcelona Supercomputing Center (Barcelona, Spain) and about four millions
of CPU hours on the FERMI cluster hosted at CINECA (Bologna, Italy). Most
of these resources were made available by a large computational program
awarded in the framework of the PRACE\footnote{http://www.prace-ri.eu}
(Partnership for Advanced Computing in Europe) initiative to enable
high-impact scientific research with a pan-European supercomputing
infrastructure, the top level of the European high performance computing
systems.

\subsection{Synthesis of X-ray emission}
\label{synthesisSNR}

From the model results, we synthesized the X-ray emission originating
from the interaction of the blast wave with the surrounding nebula,
following an approach analogous to that of previous
studies (\citealt{orlando2,2009A&A...493.1049O}). The results of numerical
simulations are the evolution of temperature, density, and velocity
of the plasma in the whole spatial domain. We rotated the system
about the three axis to fit the inclination of the ring as found
from the analysis of optical data (\citealt{2005ApJS..159...60S}), namely
$i_x = 41^{\rm o}$, $i_y = -8^{\rm o}$, and $i_z = -9^{\rm o}$.

In the case of fast collisionless shocks, as those observed in SNRs,
the synthesis of X-ray emission has to take into account the lack of
temperature-equilibration between electrons and ions. In fact, the jump
conditions at shock speeds exceeding 500 km s$^{-1}$ drive thermal
and dynamic changes of the plasma on very short timescales.  Most of
the shock energy is transferred to ions (\citealt{2007ApJ...654L..69G}),
so that the ratio of the electron to ion temperature $\beta$ in the
post-shock region is generally less than 1: the larger the shock
velocity, the smaller $\beta$. We computed this effect in our model
by considering that the equilibration in the post-shock plasma is
reached through Coulomb collisions which are expected to proceed
relatively slowly (\citealt{2007ApJ...654L..69G}), leading to $\beta < 1$.
To take into account this effect, we added an additional passive tracer to
the model equations which stores the time $t_{\rm shj}$ when the
plasma in the $j$-th domain cell was shocked. Then the electron
temperature in each cell was calculated from the time $t_{\rm shj}$
and from the ion temperature and plasma density derived by integrating
Eqs.~\ref{mod_eq}. More specifically, we assumed that the electrons
are heated almost istantaneously at the shock front up to $kT
\sim 0.3$ keV (regardless of the shock Mach number) by lower hybrid
waves (\citealt{2007ApJ...654L..69G}), as suggested for shock velocities of
the order of $10^3$ km s$^{-1}$ like those in our simulations. Then
we calculated the evolution of ion and electron temperatures in each
cell of the post-shock medium by including the effects of the Coulomb
collisions. This evolution was calculated in the time $\Delta t_{\rm j}
= t-t_{\rm shj}$ ($t$ is the current time) since when the plasma in the
cell was shocked.

Another important effect to take into account in the synthesis of
X-ray emission from fast shocks is the time lag of the plasma to
change its ionization from a cool to a hot state. If the timescale
of the temperature increase at the shock front is much shorter than
the ionization and recombination timescales, the plasma ions can
be at a lower ionization state than the equilibrium state corresponding
to the local electron temperature (\citealt{1983ApJS...51..115H}).
Such deviations from equilibrium of ionization may have dramatic
effects on the interpretation of observations. They are taken into
account in our model by following an approach discussed in the
literature (\citealt{2010MNRAS.407..812D}) which is particularly effective
in the case of 3D simulations in order to have high efficiency
together with a reasonable accuracy in the synthesis of X-ray
emission.  From the values of emission measure, electron temperature,
and maximum ionization age in the $j$-th domain cell, the corresponding
X-ray spectrum is synthesized by using the NEI emission model VPSHOCK
available in the XSPEC package along with the "NEI version 2.0"
atomic data from ATOMDB (\citealt{2001ApJ...556L..91S}). The emission
measure in the $j$-th domain cell is em$_{\rm j} = n_{\rm ej}^2
V_{\rm j}$ (where $n_{\rm ej}$ is the particle density in the cell,
$V_{\rm j}$ is the cell volume); the electron temperature in the
cell $T_{\rm ej}$ is computed by taking into account the deviations
from temperature-equilibration as described above; the maximum
ionization age in the cell is $\tau_{\rm j} = n_{\rm ej} \Delta
t_{\rm j}$ (where $\Delta t_{\rm j}$ is the time since when the
plasma in the cell was shocked; see above).

\begin{table}
\caption{Adopted parameters in the synthesis of X-ray emission}
\label{tab2}
\begin{center}
\begin{tabular}{lll}
\hline
\hline
Parameter   & Value$^a$     &  Reference  \\
\hline
$D$         &  $51.4$ kpc  &   \cite{1999IAUS..190..549P}   \\
$N_{\rm H}$ &  $2.35\times 10^{21}$ cm$^{-2}$ &   \cite{2006ApJ...646.1001P}   \\
He          &  2.57        &   \cite{2009ApJ...692.1190Z}   \\
C           &  0.03        &   \cite{2009ApJ...692.1190Z}   \\
N           &  0.56        &   \cite{2009ApJ...692.1190Z}   \\
O           &  0.081       &   \cite{2009ApJ...692.1190Z}   \\
Ne          &  0.29        &   \cite{2009ApJ...692.1190Z}   \\
Mg          &  0.28        &   \cite{2009ApJ...692.1190Z}   \\
Si          &  0.33        &   \cite{2009ApJ...692.1190Z}   \\
S           &  0.30        &   \cite{2009ApJ...692.1190Z}   \\
Ar          &  0.537       &   \cite{2009ApJ...692.1190Z}   \\
Ca          &  0.03        &   \cite{2009ApJ...692.1190Z}   \\
Fe          &  0.19        &   \cite{2009ApJ...692.1190Z}   \\
Ni          &  0.07        &   \cite{2009ApJ...692.1190Z}   \\
\hline
\end{tabular}\\
\flushleft{$^a$ The abundances are relative to the solar photospheric
values (\citealt{1989GeCoA..53..197A}).}
\end{center}
\end{table}

We assumed the source at a distance $D = 51.4$ kpc
(\citealt{1999IAUS..190..549P}). We adopted the metal abundances derived
from the analysis of deep {\it Chandra}/LETG and HETG observations
of SN\,1987A (\citealt{2009ApJ...692.1190Z}) and summarized in
Table~\ref{tab2}. The X-ray spectrum from each cell was filtered through
the photoelectric absorption by the ISM, assuming a column density $N_{\rm
H} = 2.35\times 10^{21}$ cm$^{-2}$ (\citealt{2006ApJ...646.1001P}). We
integrated the absorbed X-ray spectra from the cells in the whole
spatial domain. The spectra are then folded through the instrumental
response of the X-ray instruments of interest, obtaining the relevant
focal-plane spectra. In such a way, we derived X-ray spectra as they
would be collected with {\it XMM-Newton}/EPIC and, in the near future,
with {\it Athena}/WFI, and X-ray images as they would be collected with
{\it Chandra}/ACIS.  The synthesized spectra and images were put in a
format identical to that of real X-ray observations and analyzed with
the standard data analysis system used for the specific instruments
of interest.

\section{Results}
\label{sec3}

\subsection{Post-explosion evolution of the supernova}
\label{evolSN}

\begin{figure}[!t]
  \centering
  \includegraphics[width=8.5cm]{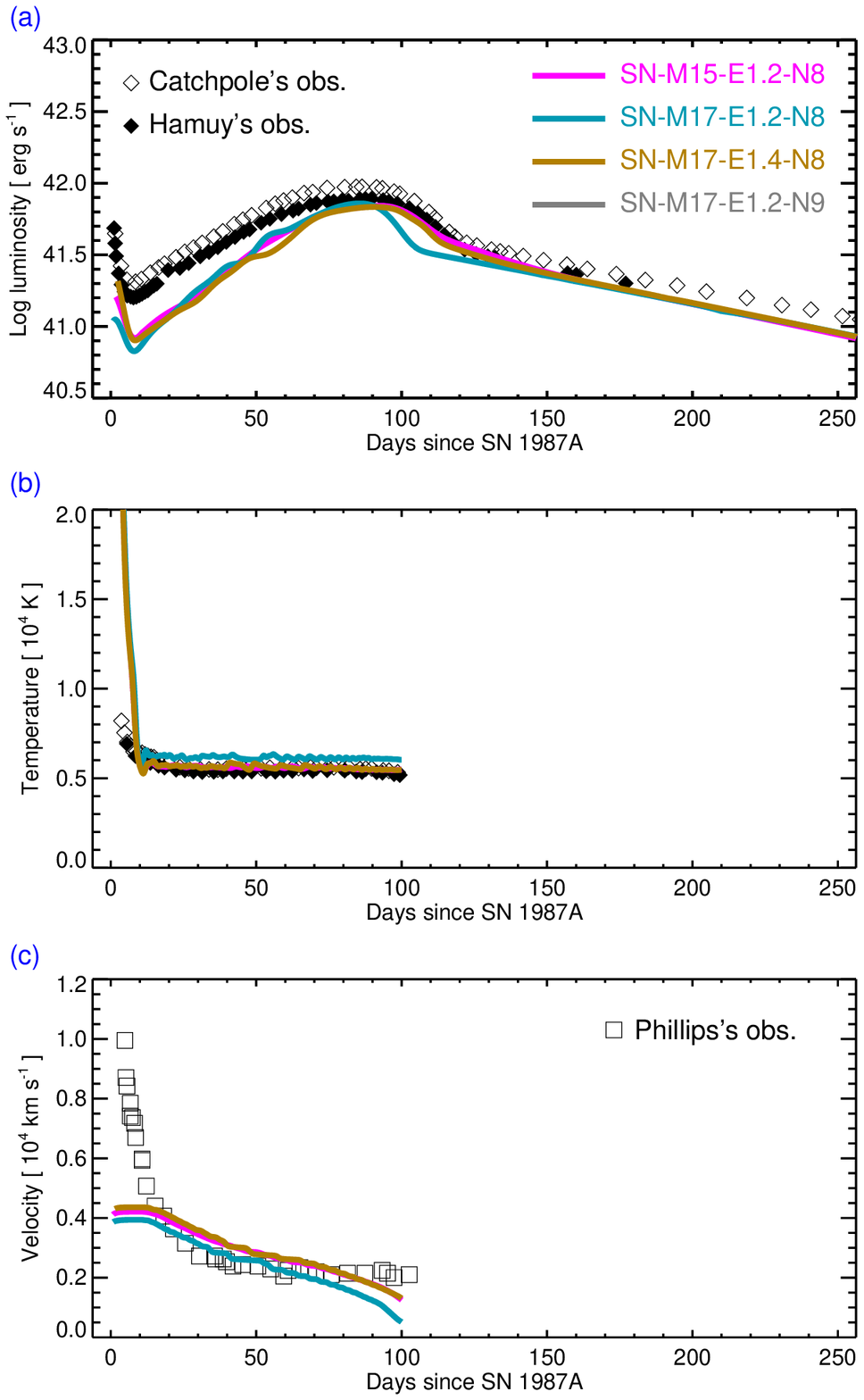}
  \caption{Bolometric lightcurves (a) and evolution of
  the photospheric temperature (b) and velocity (c) derived from the
  models listed in Table~\ref{tabSN} (solid colored lines) compared to
  the corresponding quantities of SN\,1987A (\citealt{1987MNRAS.229P..15C,
  1988MNRAS.231P..75C, 1988AJ.....95...63H, 1988AJ.....95.1087P})
  (empty squares and filled and empty diamonds).  The photospheric
  temperature and velocity can be defined from the model during the
  so-called photospheric phase (see text), corresponding to the first
  $\sim 100$ days of evolution.  Models
  SN-M17-E1.2-N8 and SN-M17-E1.2-N9 overlap each other.}
\label{lc_bolom} \end{figure}

The post-explosion evolution of the simulated SN\,1987A follows the
trend described in details by \cite{2011ApJ...741...41P}. Here
we summarize the main phases. The evolution is guided by the
thermodynamics of the expanding ejecta and passes through three different
phases. Initially the envelope is completely ionized and optically thick,
and the emission is due to the release of internal energy on a diffusion
timescale (diffusive phase). Then the ejecta recombine and the emission is
dominated by the sudden release of energy caused by the receding motion
of the wavefront through the envelope (recombination phase).  Finally,
the envelope is recombined and optically thin to optical photons, and
the emission comes from the thermalization of the energy deposited by
gamma-ray photons (radioactive-decay phase or nebular phase). Usually
the first two phases are globally referred as photospheric phase, during
which it is possible to compare the observed photospheric temperature
and velocity with the corresponding simulated quantities, contrarily to
the nebular phase where the same notion of photosphere loses its meaning
(\citealt{2011ApJ...741...41P, 2011ApJ...729...61B}). We adopted the
so-called inflection time $t_{\rm inf}$ (measured from the explosion
epoch) as a measurement of the duration of the photospheric phase
(\citealt{2013MNRAS.434.3445P}); for SN\,1987A $t_{\rm inf} \approx
100$ days.

\begin{figure}[!t]
  \centering
  \includegraphics[width=8.5cm]{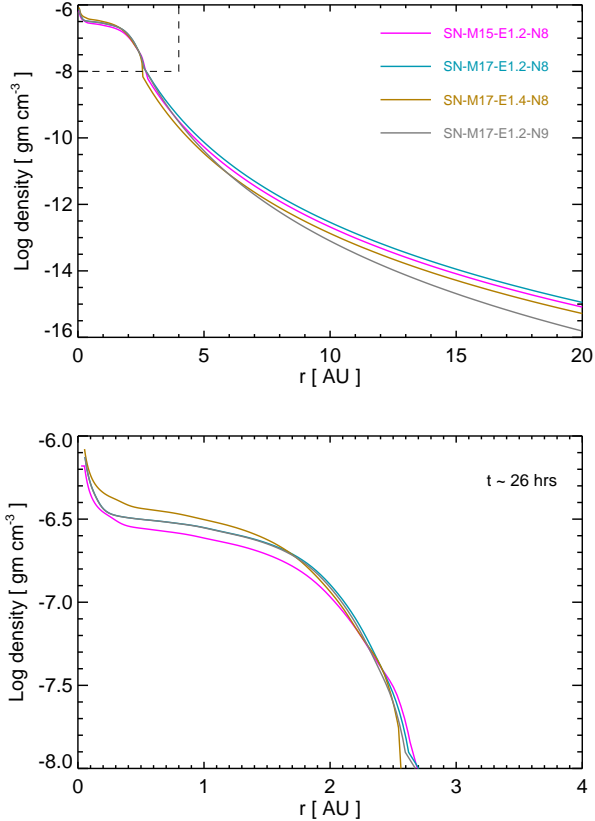}
  \caption{(Top) Radial density profiles of ejecta for
  the models reproducing the main observables of SN\,1987A (see
  Table~\ref{tabSN} and Fig.~\ref{lc_bolom}) at $t\sim 26$ hours
  after the shock breakout. Different colors mark different models.
  (Bottom) Enlargment of the region marked with a box in the top
  panel, showing the details of the bulk of the ejecta.}
\label{ejecta_prof}
\end{figure}

\begin{table}
\caption{Parameters of the SN models reproducing the main observables
of SN\,1987A during the first 250 days from the outburst.}
\label{tabSN}
\begin{center}
\begin{tabular}{lccccc}
\hline
\hline
Model           &  $M_{\rm env}$  & $E_{\rm SN}$     &  $\alpha$ & $R_0$  & $M_{Ni}$  \\
                &  [$M_{\odot}$]  & [$10^{51}$~erg]  &           & [$10^{12}$~cm] &
[$M_{\odot}$]  \\
\hline
SN-M17-E1.2-N8  &  17             & 1.2              &  -8  & 3 & 0.07  \\
SN-M15-E1.2-N8  &  15             & 1.2              &  -8  & 3 & 0.07  \\
SN-M17-E1.4-N8  &  17             & 1.4              &  -8  & 3 & 0.07  \\
SN-M17-E1.2-N9  &  17             & 1.2              &  -9  & 3 & 0.07  \\
\hline
\end{tabular}
\end{center}
\end{table}

From the models, we derived the main observables (namely the bolometric
lightcurve and the evolution of the photospheric temperature and velocity)
during the first 250 days after explosion. Then we compared the model
results with observations of SN\,1987A by performing a simultaneous
$\chi^2$ fit of these observables, thus constraining the ejected
mass and the explosion energy of the SN. We found the comparison with
the observations satisfactory for models with a total energy ranging
between 1.2 and $1.4\times 10^{51}$~erg and an envelope mass between
15 and 17~$M_{\odot}$ (see Fig.~\ref{lc_bolom}). Figure~\ref{ejecta_prof}
shows the radial distribution of mass density 26 hours after the
breakout of the shock wave at the stellar surface for the models best
reproducing the observations (see Table~\ref{tabSN} for details). Note
that runs SN-M17-E1.2-N8 and SN-M17-E1.2-N9 differ only for the slope
of the power-law describing the high-velocity shell of the SN (see
Section~\ref{evolSN}): the density and kinetic energy in the outer
envelope are lower for the steeper profile of density. The ejecta
distribution in this outer shell does not change significantly the
observables of the SN.

The difficulty of reproducing the observables at early time in
Fig.~\ref{lc_bolom} is mainly due to the initial conditions used in
our simulations (see Section~\ref{modelSN}), while some less prominent
discrepancies at late times ($80-110$ days) might be explained by
the absence of non-thermal ionization from gamma rays in our model
(\citealt{2011ApJ...741...41P}).

The comparison of model results with observations enabled us to constrain
the bulk of envelope mass $M_{\rm env}$ and the enegy of the explosion
$E_{\rm SN}$. On the other hand, no firm conclusions on the distribution
of energy and mass in the high-velocity shell of the SN were obtained
from the analysis of SN observables.

\subsection{Remnant expansion through the circumstellar nebula}
\label{evolSNR}

\begin{figure*}[!t]
  \centering
  \includegraphics[width=15cm]{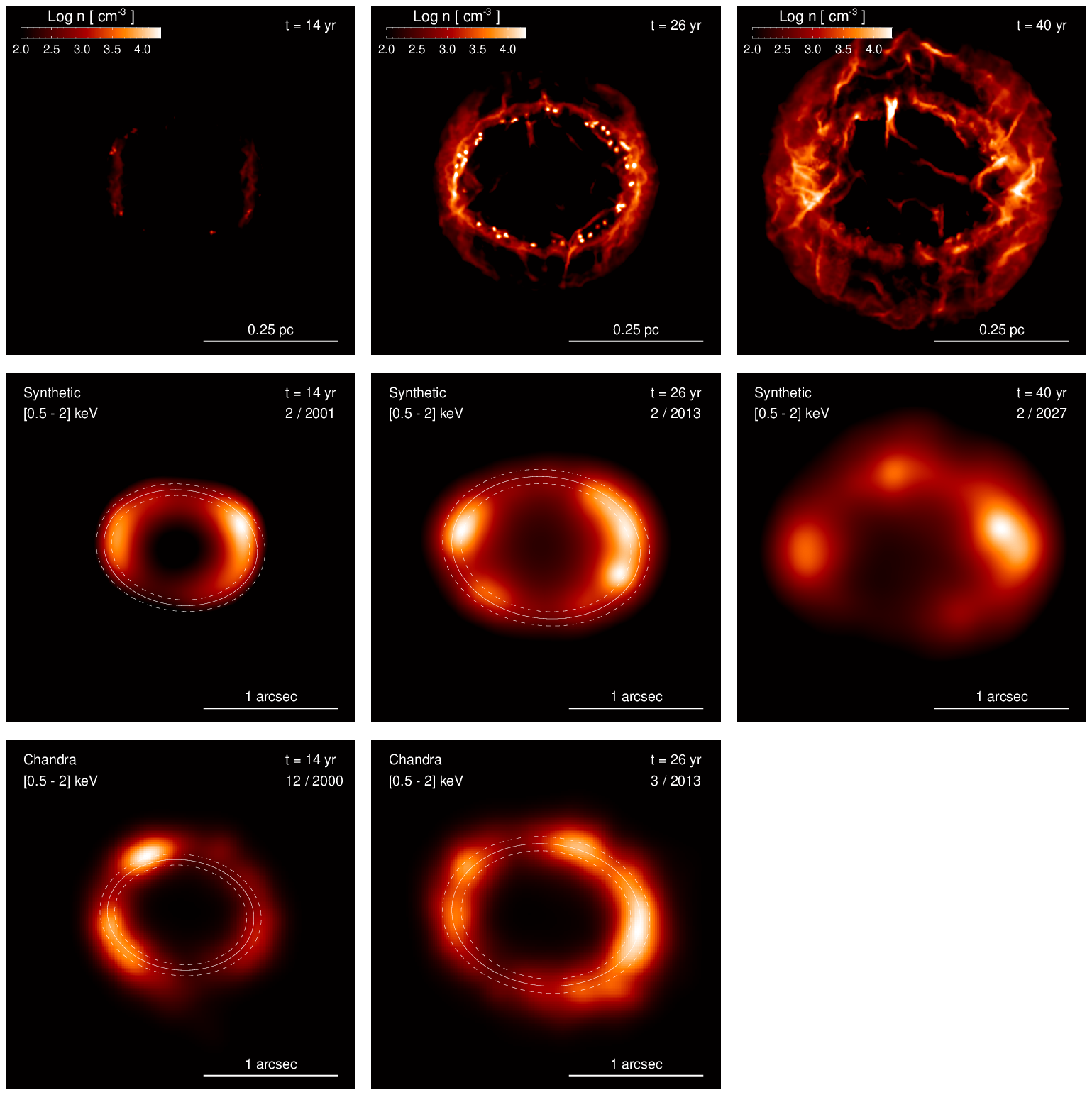}
  \caption{Interaction of the blast wave with the nebula.
  (Top) Three-dimensional volume rendering of particle density of
  the shocked plasma at the labeled times. (Middle) Corresponding
  synthetic maps of X-ray emission in the $[0.5,2]$~keV band
  integrated along the line-of-sight. Each image has been normalized
  to its maximum for visibility and convolved with a Gaussian of
  size 0.15 arcsec to approximate the spatial resolution of
  {\it Chandra} observations (\citealt{2013ApJ...764...11H}).
  (Bottom) Maps of X-ray emission of SN\,1987A collected with
  {\it Chandra} at the labeled times, and normalized to their
  maximum for visibility (see Appendix~\ref{datanalysis}). The overplotted
  ellipsoids represent the projection of circles lying in the equatorial
  plane of SN\,1987A and fitting the position of the maximum X-ray
  emission in each observation. The dashed lines show an uncertainty of
  10\%. (See on-line movie).}
\label{fig2}
\end{figure*}

\begin{figure}[!t]
  \centering
  \includegraphics[width=8.5cm]{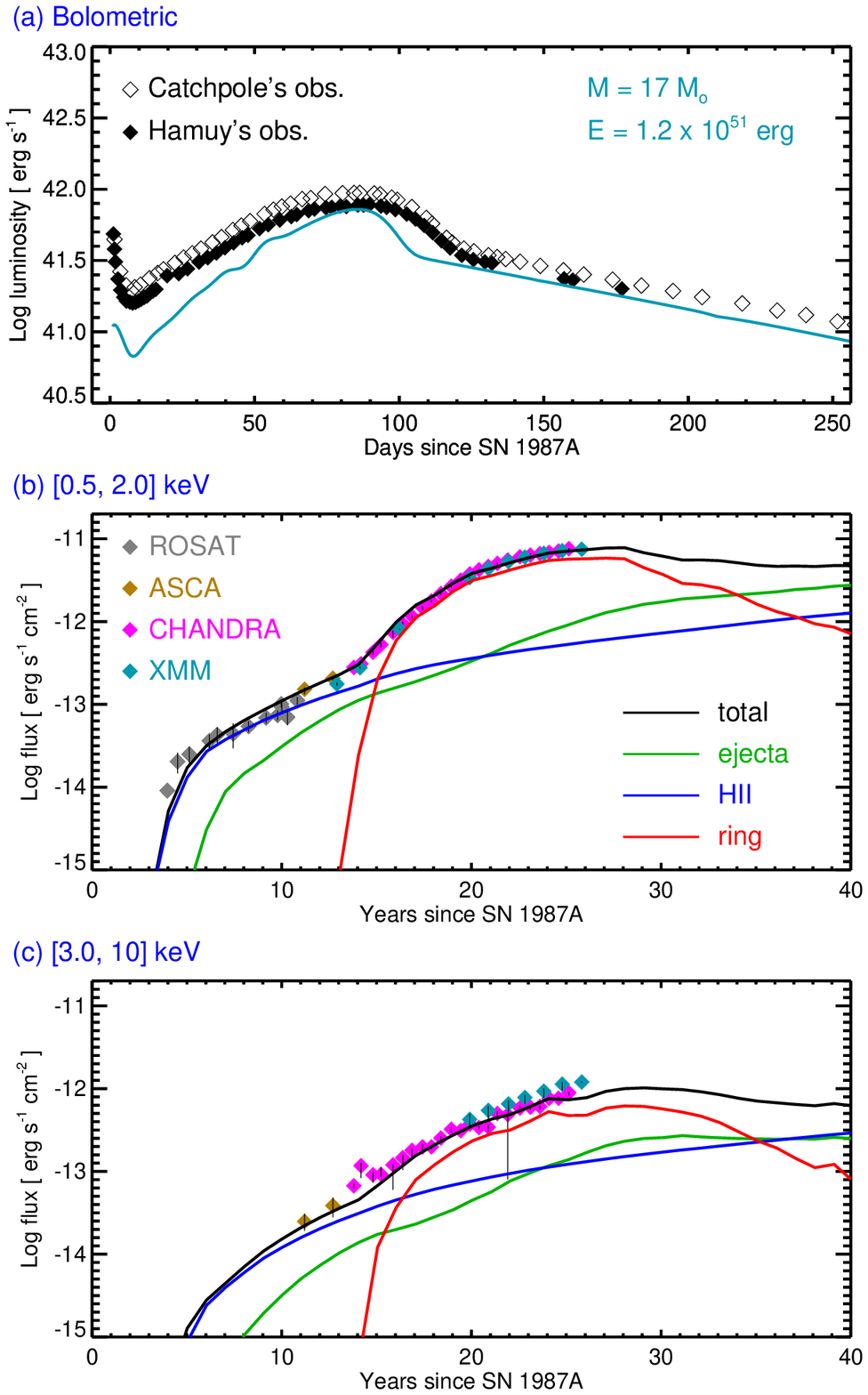}
  \caption{Observed and modeled lightcurves. (a) Bolometric
  lightcurve of our favored model (cyan line) compared
  to the lightcurve of SN\,1987A (filled and empty diamonds;
  \citealt{1987MNRAS.229P..15C,1988MNRAS.231P..75C,1988AJ.....95...63H}).
  (b) X-ray lightcurve in the $[0.5, 2]$~keV band synthesized from the
  favored model (black line) compared to the lightcurve of SN\,1987A
  observed with Rosat (grey diamonds; \citealt{2006A&A...460..811H}),
  {\it ASCA} (brown; see Appendix~\ref{datanalysis}), {\it Chandra}
  (magenta; \citealt{2013ApJ...764...11H}) and {\it XMM-Newton}
  (cyan; \citealt{2006A&A...460..811H, 2012A&A...548L...3M};
  see Appendix~\ref{datanalysis}).  Green, blue and red lines mark
  the contribution to emission from the shocked ejecta, the shocked
  plasma from the H\,II region, and the shocked plasma from the ring,
  respectively. (c) Same as in Fig.~\ref{fig1}b but for the lightcurve
  in the $[3, 10]$~keV band.}
\label{fig1} \end{figure}

We followed the evolution from the SN phase to the SNR phase for
40 years, restricting our analysis to SN models reproducing the main
observables of SN\,1987A during the first 250 days of evolution
(listed in Table~\ref{tabSN}). Then, we explored the space of
parameters describing the CSM (mainly the equatorial ring and the
H\,II region) and compared the X-ray lightcurves and spectra
synthesized from the models with those observed in SN\,1987A (see
Fig~\ref{fig1}).

\begin{figure}[!t]
  \centering
  \includegraphics[width=8.cm]{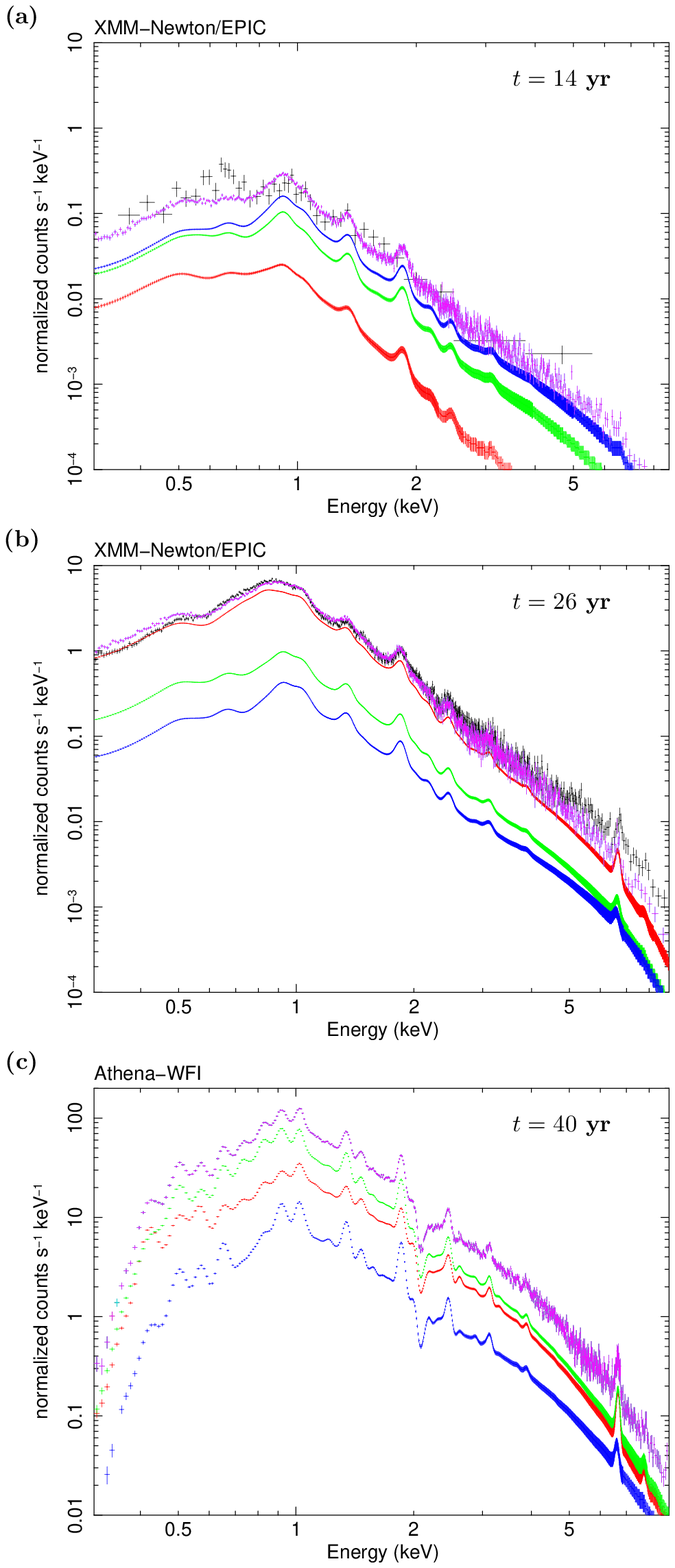}
  \caption{Synthetic and observed X-ray spectra of SN\,1987A. (a)
  {\it XMM-Newton}/EPIC-pn spectra at $t = 14$~yr. The true spectrum
  is marked in black (see Appendix~\ref{datanalysis}); the synthetic
  spectrum from the whole shocked plasma is marked in magenta; the
  contributions to emission from the different shocked plasma components
  are marked in green (ejecta), red (ring), and blue (H\,II region). (b)
  As in Fig.~\ref{fig3}a, for $t=26$~yr.  (c) As in Fig.~\ref{fig3}a,
  for $t=40$~yr and for spectra as they would be
  collected with {\it Athena}/WFI.}
\label{fig3} \end{figure}

\begin{figure*}[!t]
  \centering
  \includegraphics[width=13.cm]{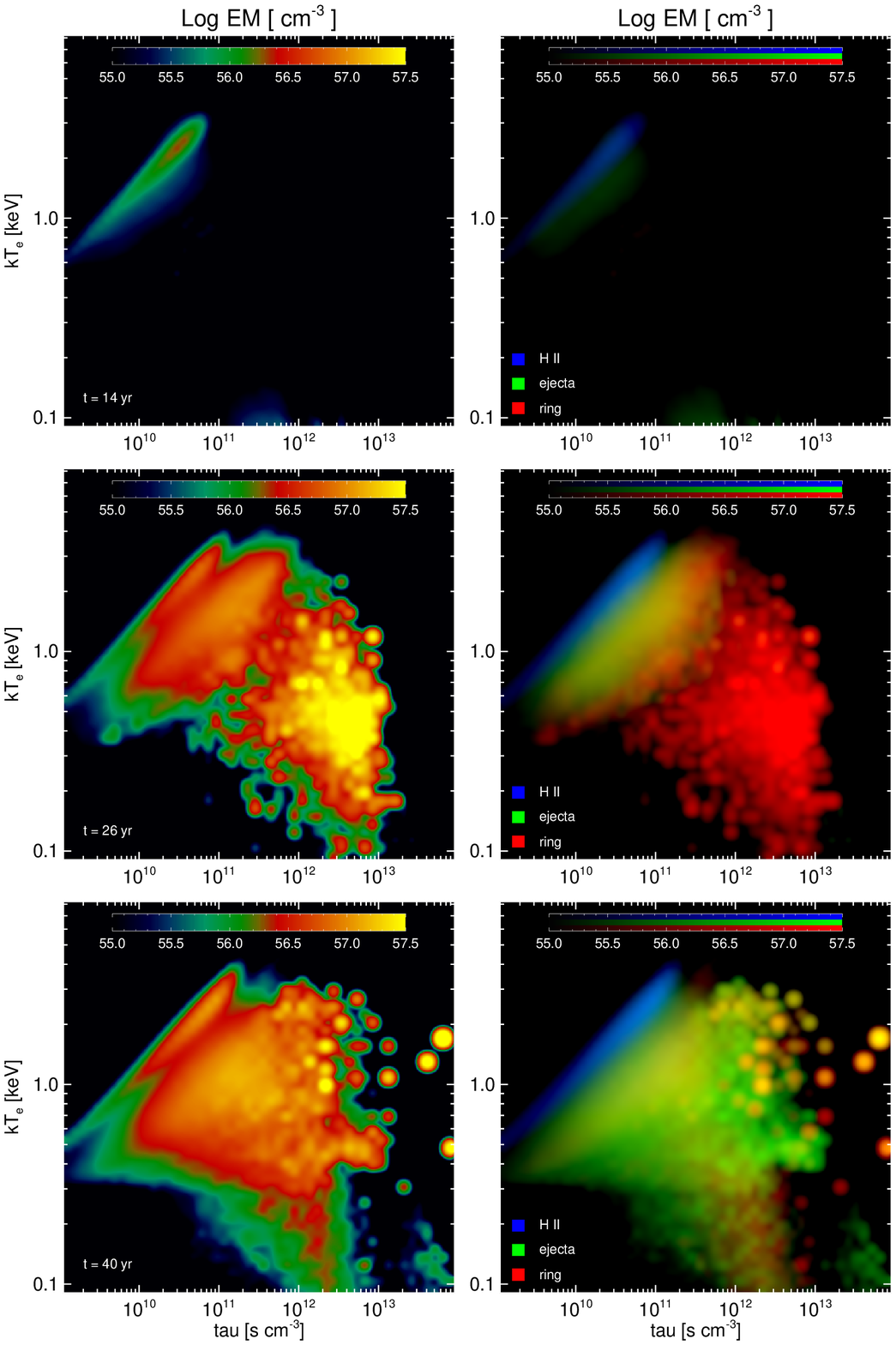}
  \caption{Left: distributions of emission measure vs. electron
  temperature $kT_{\rm e}$ and ionization time scale $\tau$ at the
  labeled times, corresponding to the X-ray spectra shown in
  Fig.~\ref{fig3}. Right: corresponding three-color composite images
  of the emission measure distributions. The colors show the contribution
  to emission measure from the different shocked plasma components,
  namely the ejecta (green), the ring (red), and the H\,II region
  (blue).}
\label{maps} \end{figure*}

\begin{figure*}[!t]
  \centering
  \includegraphics[width=16.cm]{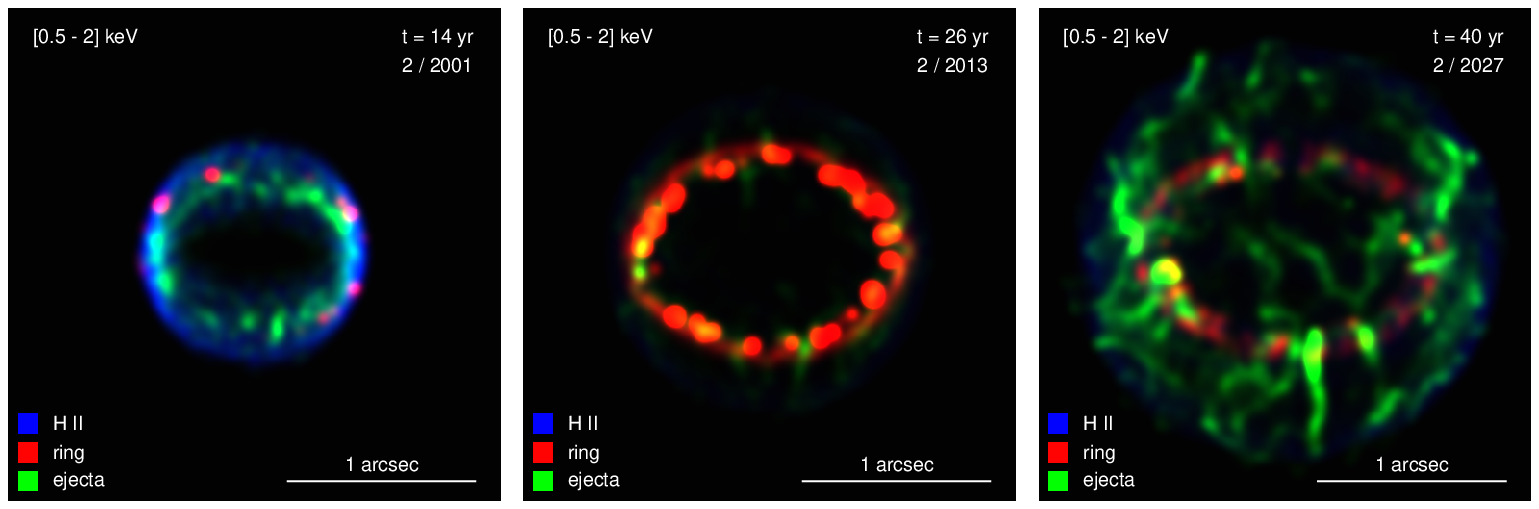}
  \caption{Three-color composite images of the X-ray emission in
  the $[0.5,2]$~keV band integrated along the line-of-sight at the
  labeled times. Each image has been normalized to its maximum for
  visibility and smoothed with a Gaussian of size 0.025 arcsec. The
  colors in the composite show the contribution to emission from
  the different shocked plasma components, namely the ejecta (green),
  the ring (red), and the H\,II region (blue). (See the on-line movie).}
\label{maps_mc} \end{figure*}

In all the cases investigated, the evolution follows the same general
trend. Initially the blast wave from the SN propagates through the wind
of the progenitor BSG. About 3 years after explosion, the ejecta reach the
H\,II region (see on-line movie) and the transition from SN to SNR enters
into its first phase of evolution (H\,II-region-dominated phase). A
forward and a reverse shocks are generated, the former propagating
into the H\,II region and the latter driven back into the ejecta (see
Fig.~\ref{fig2}). The X-ray emission begins to increase rapidly with
time and is dominated by emission from shocked plasma of the H\,II
region and by a smaller contribution of shocked ejecta in the outer SN
envelope. This is evident from the lightcurves (Fig.~\ref{fig1}), the
spectra (Fig.~\ref{fig3}), and the emission morphology (Fig.~\ref{maps_mc}
and on-line movie). The emitting plasma is largely out of equilibrium
of ionization and its emission measure distribution as a function of
electron temperature, $kT_{\rm e}$, and ionization time scale, $\tau$,
is peaked at $kT_{\rm e} \approx 2$~keV and $\tau\approx 3\times 10^{10}$
s cm$^{-3}$ with a sharp distribution that can be approximated with an
isothermal plasma component (see upper panels of Fig.~\ref{maps}). This
is in excellent agreement with the best-fit parameters derived from the
spectral fitting with isothermal components of X-ray spectra of SN\,1987A
at this epoch (see Appendix~\ref{datanalysis}). This phase lasts until
the blast wave hits the equatorial ring (year 15).

\begin{figure}[!t]
  \centering
  \includegraphics[width=8.5cm]{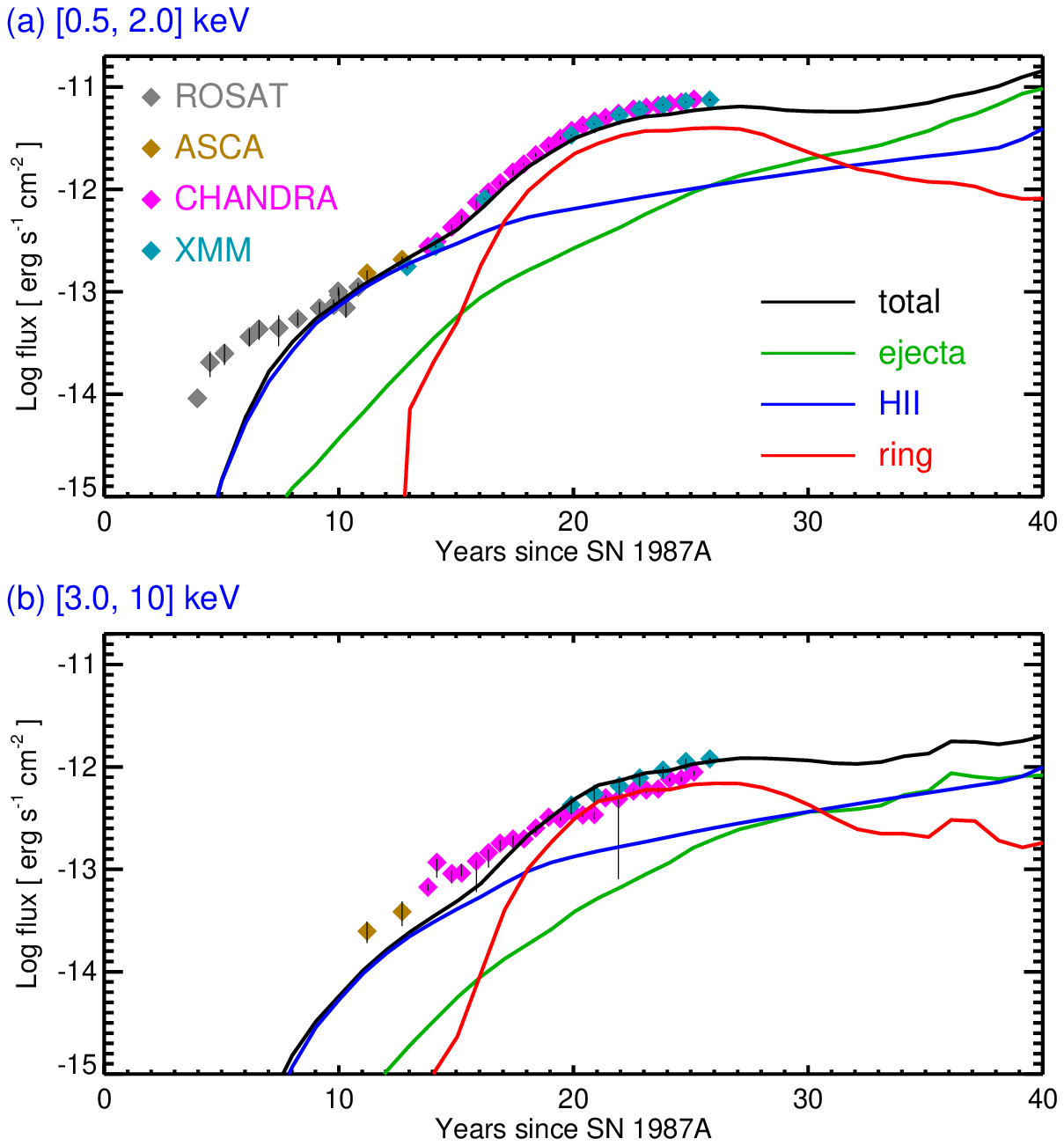}
  \caption{As in Fig.~\ref{fig1}b and \ref{fig1}c for a model with envelope mass
  $M_{\rm env} = 17 M_{\odot}$, ejecta energy $E_{\rm SN} = 1.2\times
  10^{51}$~erg, and with the density profile of ejecta in the
  high-velocity shell approximated by a power-law with index $\alpha =
  -9$ (run SN-M17-E1.2-N9 in Table~\ref{tabSN}). The parameters of the
  CSM of this model are reported in Table~\ref{modl9}.}
\label{prof_n9}
\end{figure}

In the H\,II-region-dominated phase, we found that the sudden increase
observed in the soft X-ray band ($[0.5,2]$~keV) is best-fitted by
models in which the outer layers of ejecta have a post-explosion radial
profile of density approximated by a power law with index $\alpha = -8$
(see Fig.~\ref{fig1}b). On the contrary, we found that models with index
$\alpha = -9$ systematically underestimate the soft X-ray flux during the
first 7 years even by changing the values of $n_{\rm HII}$ and $r_{\rm
HII}$ within the range of values compatible with observations. This is
evident in Fig.~\ref{prof_n9} where we report the result for the model
with $\alpha = -9$ best approximating the observations: the discrepancy
between model results and observations is evident for $t < 7$ yr even
though the same model fits well the bolometric lightcurve of the SN during
the first 250 days of evolution (see Fig.~\ref{lc_bolom}) and the X-ray
lightcurves for $t>7$~years (Fig.~\ref{prof_n9}). Table~\ref{modl9}
reports the parameters of this model to be compared with those of the
model with $\alpha = -8$ shown in Table~\ref{tab1}. The reason for this
discrepancy is the content of energy and mass of the outer shell of ejecta
which is the first material hitting the nebula and, thus, determining
the early X-ray emission from the ejecta-nebula interaction. Models
with $\alpha = -9$ have less mass and energy in the outer shell than
models with $\alpha = -8$ and, as a result, the corresponding soft X-ray
lightcurve rises more slowly and fails in fitting the observations.

\begin{table}
\caption{Parameters of the CSM for the hydrodynamic model with $\alpha = -9$
best approximating the X-ray lightcurves}
\label{modl9}
\begin{center}
\begin{tabular}{lclc}
\hline
\hline
CSM component & Parameters     & Units & Best-fit values  \\
\hline
BSG wind: & $\dot{M}_{\rm w}$  & ($M_\odot$~year$^{-1}$) &  $10^{-7}$     \\
  & $v_{\rm w}$    &  (km~s$^{-1}$)   &   500    \\
  & $r_{\rm w}$    &  (pc)            &   0.05   \\
\hline
H\,II region: & $n_{\rm HII}$  &  ($10^2$ cm$^{-3}$) &  20     \\
 & $r_{\rm HII}$  &  (pc)            &  0.1    \\
\hline
Equatorial ring: & $n_{\rm rg}$   &  ($10^3$ cm$^{-3}$) &  1  \\
 & $r_{\rm rg}$   &  (pc)            &   0.18   \\
 & $w_{\rm rg}$   &  ($10^{17}$ cm)  &  $1.7$   \\
 & $h_{\rm rg}$   &  ($10^{16}$ cm)  &  $3.5$   \\
\hline
Clumps: & $<n_{\rm cl}>$ &  ($10^4$ cm$^{-3}$) &  $1.3\pm 0.3$      \\
 & $<r_{\rm cl}>$ &  (pc)            &  $0.17\pm 0.015$     \\
 & $w_{\rm cl}$   &  ($10^{16}$ cm)  &  $1.7$           \\
 & $N_{\rm cl}$   &                  &  40         \\
\hline
\end{tabular}
\end{center}
\end{table}

Note that the value of $\alpha$ have little (if any) effect in
determining the main observables of the SN (e.g. the bolometric
lightcurve) during the first 250 days of evolution. SN models with
the same envelope mass and total energy but with different values
of $\alpha$ lead to very similar results (see Fig.~\ref{lc_bolom}).
This is due to the fact that the observables of the SN depend on
the bulk of ejecta. Our findings, therefore, shows that the X-ray
emission originating from the SNR in this phase can constrain the
structure of outermost ejecta better than the emission from the SN.
Since the density profile of ejecta is expected to depend on the
structure of the progenitor star and on the shock acceleration of
the gas during the explosion, studying the early interaction of the
ejecta with the nebula may provide important clues to the latest
stage of stellar evolution and may be used as a probe of the
mechanisms involved in the SN engine.

During the H\,II-region-dominated phase, the model enabled us to
constrain also the parameters characterizing the H\,II-region (namely
its density and inner radius); the best-fit parameters are listed
in Table~\ref{tab1}.

\begin{figure}[!t]
  \centering
  \includegraphics[width=8.5cm]{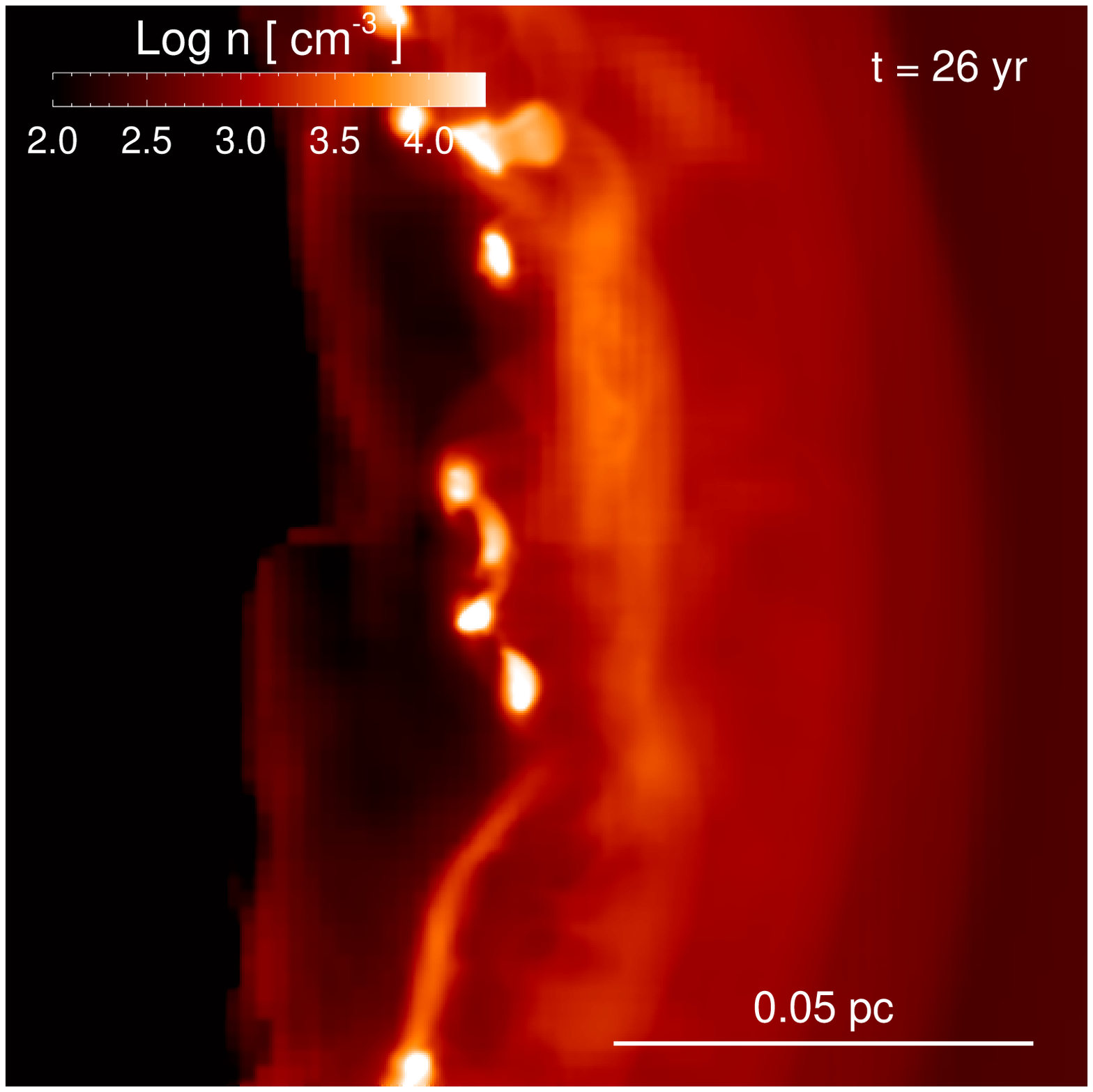}
  \caption{Three-dimensional volume rendering of particle density of the
  shocked plasma at $t=26$~yr. The line-of-sight is aligned with the $z$
  axis and the supernova is located to the left of this plot. The figure
  shows a close-up view of the interaction of the ejecta with the ring
  in the equatorial plane. The bright knots are shocked clumps of the
  ring. (See the on-line movie).}
\label{figclumps}
\end{figure}

Around year 13 the blast wave hits the first dense clump of the
equatorial ring and around year 15 it reaches the inner edge of the
ring (see Fig.~\ref{fig2} and on-line movie). The transition from SN
to SNR enters into the second phase of evolution (ring-dominated phase)
which lasts until year 32, when the forward shock propagates beyond the
majority of the dense ring material. This phase is characterized by the
interaction of the forward shock with the dense clumps of the ring. Each
shocked clump evolves toward a core-plume structure with a crescent-like
shape characterized by Kelvin-Helmholtz instabilities developing in
the downstream region (see Fig.~\ref{figclumps} and on-line movie).
As the shock travels through the clump, Rayleigh-Taylor instabilities
develop on the upstream side of the clump, leading to its progressive
fragmentation. A complex pattern of filaments and fragments forms in the
interclump region, with densities varying between $\sim 10^3$~cm$^{-3}$
in the smooth component of the ring and $\sim 10^4$~cm$^{-3}$ in
proximity of the clumps. Note that the appropriate description of the
shock-clump interaction required the high-spatial resolution adopted in
our simulations. The modelled features cannot be reproduced at lower
resolutions (e.g. \citealt{2014ApJ...794..174P}).

In this phase, the contribution from the shocked ring dominates the X-ray
emission (see Figs.~\ref{fig1}-\ref{maps_mc} and on-line
movie). The shocked cores of the clumps lead predominantly to soft
X-rays and determine the further steepening of the soft X-ray lightcurve
(Fig.~\ref{fig1}b). This emitting plasma component is essentially in
collisional ionization equilibrium and its emission measure peaks to
electron temperature $kT_{\rm e}\approx 0.5$~keV and ionization time
scale $\tau\approx 5\times 10^{13}$ s cm$^{-3}$ (see middle panels of
Fig.~\ref{maps}). We can roughly identify this material with the plasma
component with $\tau > 10^{13}$ s cm$^{-3}$ derived from the spectral
fitting of X-ray spectra of SN\,1987A collected with current X-ray
observatories for $t > 15$~yr (e.g. \citealt{2013ApJ...764...11H}; see
also Appendix~\ref{datanalysis}). The smooth component of the ring and the
fragments of the shocked clumps stripped by hydrodynamic instabilities
dominante the emission in the hard band. This plasma component causes
the broadening of the emission measure distribution around the peak
due to shocked clumps and includes plasma with $kT_{\rm e}$ up to $\sim
2$~keV and $\tau$ down to $\sim 10^{11}$ s cm$^{-3}$ (see middle panels
of Fig.~\ref{maps}). Also a significant contribution to the emission
with $kT_{\rm e} > 1$~keV and $\tau < 10^{11}$ s cm$^{-3}$ comes from the
shocked ejecta. Given the complexity of the emission measure distribution
in this phase, it is not obvious to attribute a physical meaning to the
isothermal components with $\tau < 10^{13}$ s cm$^{-3}$ derived from the
spectral fitting of the X-ray spectra of SN\,1987A and largely used in
the literature (see Appendix~\ref{datanalysis}). Indeed the physical
origin of the observed X-ray spectra is unveiled by our hydrodynamic
model, as shown in Fig.~\ref{maps}. From the model best reproducing the
X-ray lightcurves and spectra, we were able to constrain the parameters
characterizing the equatorial ring (see Table~\ref{tab1} for details).

During the first two phases of evolution, the remnant morphology in
the X-ray band synthesized from the model appears very similar to that
observed. In particular, in the ring-dominated phase, the morphology is
characterized by bright knots originating from the shocked clumps and
resambles that of SN\,1987A (see Figs.~\ref{fig2}, \ref{maps_mc}, and
on-line movie). In Fig.~\ref{fig2} we compare the synthetic and the observed
morphology at two different epochs (years 14 and 26). The ellipsoids
overplotted in the figure represent the projection of circles lying in
the equatorial plane of SN\,1987A and fitting the position of the maximum
X-ray emission in the observations. Although the extension of the X-ray
source synthesized from the model seems to be smaller than the observed
one (suggesting a modelled blast wave slightly slower than observed),
the synthetic maps fit those observed within an uncertainty of 10\%
(dashed lines). In particular, at year 14, the observations show a bright
knot at north-west that may indicate that the observed blast wave is at
a distance larger than that in our model. On the other hand, the knot
is also well beyond the ellipsoid fitting the position of the forward
shock in the equtorial plane, suggesting that, probably, the knot is the
result of the interaction of the blast wave with some inhomogeneity at
some height above the equatorial plane. This feature could be reproduced
in our model considering, for instance, an overdense clump located well
above the equatorial plane.

The remnant enters into the third phase (ejecta-dominated
phase) around year 32, when the contribution of shocked ejecta
to the soft X-ray emission becomes the dominant component (see
Figs.~\ref{fig1}-\ref{maps_mc}). The reverse shock travels through
innermost ejecta with higher densities. Now the emitting plasma is
characterized by a broad emission measure distribution that peaks
at $kT_{\rm e} \approx 1$~keV and $\tau \approx 5\times 10^{11}$ s
cm$^{-3}$ although it is also characterized by few spikes with $\tau >
10^{13}$ s cm$^{-3}$ due to the interaction of high-density clumps of
ejecta with the ring (see lower panels of Fig.~\ref{maps} and right panel
in Fig.~\ref{maps_mc}). The remnant morphology shows a revival of very
bright knots but now due to shocked clumps of ejecta rather than clumps
of the ring (see Fig.~\ref{fig2} and on-line movie). Our simulations
show that, in this phase, the evolution of the X-ray emission depends
again on the density distribution of the outermost ejecta as in the
H\,II-region-dominated phase. Models with the same explosion energy
and envelope mass but with a different slope of the density profile of
ejecta lead to significantly different X-ray lightcurves after year 32
(compare Fig.~\ref{fig1} and Fig. \ref{prof_n9}). In particular we found
that the steeper is the density profile, the higher is the contribution
of shocked ejecta to X-ray emission after year 32. In the next future,
therefore, SN\,1987A will offer the possibility to study directly the
structure and chemical composition of innermost ejecta and the imprint
of the metal-rich layers inside the progenitor star.

\subsection{The central compact remnant}

Pioneering studies by \cite{1987ApJ...322L..15F} suggested
that the dense shells of expanding ejecta may obscure the X-ray
emission of the compact object (if any) of SN\,1987A for a few
decades. Our model describes the distribution of the ejecta (shocked
and unshocked) as well as the complex circumstellar environment.
Thus the model allows us to get a thorough description of the local
absorption in the remnant and to derive constraints on the X-ray
emission of its yet undetected central source.

The effective surface temperature of an isolated neutron
star with an age of $10-30$~yr is expected to be in the range $T_{\rm
eff} \sim 3-6$~MK (e. g., \citealt{2008AstL...34..675S}). This
temperature should stay constant until the age of the star is lower
than its thermal relaxation time, $t_{\rm rel}$, which is in the
range $30-300$~yr, depending on the properties of the stellar crust
and of its core (e.g. \citealt{2008AstL...34..675S}). From the
analysis of {\it Chandra} observations, \cite{2009ApJ...706L.100N}
derived an upper limit of $0.010$ counts per second for the central
source (in the $[0.08, 10]$~keV energy band) on the basis of its
non-detection in the {\it Chandra}-HRC data. Without accounting for
the local absorption, this upper limit corresponds to a surface
temperature lower than $\approx 2.5$~MK, which is at odds with
predictions from standard theories and may suggest extremely short
values of $t_{\rm rel}$ due, for example, to superfluidity or quark
matter.

Our model shows that, at the present epoch, the column
density toward the center of the remnant\footnote{We inspected a
region $\approx 3\times 10^{17}$~cm wide (corresponding to $\approx
0.2$~arcsec considering the distance of $\approx 51.4$~pc) to account
for a possible proper motion of the central object around the
geometrical center of the remnant, finding limited variations of
the absorbing column.} is $N_{\rm H} \approx 4.5-5\times10^{22}$~cm$^{-2}$,
mainly associated with the unshocked ejecta. This local absorbing
column density is almost a factor of 20 higher than the interstellar
absorption toward the Large Magellanic Cloud and allows us to revise the
constraints on the thermal emission from the central source suggested
by \cite{2009ApJ...706L.100N}.  By accounting for the local absorption,
we then obtained an upper limit of $3.9$~MK for the surface temperature
(i.e. a bolometric luminosity $L_{\rm bol}=1.6\times
10^{35}$~erg~s$^{-1}$), by considering a neutron star with radius
$R_{\rm NS}=10$~km at a distance of $51.4$~kpc. This revised value
is in good agreement with standard theories, which suggest bolometric
luminosities $L_{\rm bol} >5\times 10^{34}$~erg~s$^{-1}$ (e.g.
\citealt{2004ARA&A..42..169Y}), without invoking the presence of
very short values of $t_{\rm rel}$.

The X-ray emission from young pulsars is typically dominated
by a nonthermal component. Therefore, we also considered the case
of nonthermal radiation, by assuming a characteristic power-law
index $\Gamma=1.5$. Taking into account the high local absorption
from the unshocked ejecta, we found that the count-rate upper limit
obtained with {\it Chandra} converts to a luminosity $L_{\rm nt}\sim
6\times 10^{35}$ erg s$^{-1}$ in the $[2, 10]$ keV band, corresponding
to a flux of $\sim 0.1$ mCrab at $51.4$ kpc (to be compared with
$L_{\rm nt}\sim 7\times 10^{34}$~erg~s$^{-1}$ obtained by
\citealt{2009ApJ...706L.100N}).

Finally we studied the variation of the local absorption with
time, finding a relatively fast decrease. In particular, we verified
that the expansion of the ejecta will reduce the column density by
a factor of $\sim 2$ ($N_{\rm H} \approx 2\times 10^{22}$~cm$^{-2}$)
at the end of our simulation (corresponding to 40 yr after the
explosion), namely at the presumed launch date of the {\it Athena}
X-ray observatory (\citealt{2013arXiv1306.2307N}).

It is worth noting that, according to our model, the bulk
of the absorption originates in the ejecta. The high metallicity
of the expanding ejecta strongly enhances their optical depth with
respect to a medium with a solar chemical composition, the O rich
ejecta heavily absorbing the X-ray radiation below 1 keV, and the
Si, S, and Fe-rich ejecta contributing at higher energies (see
\citealt{2000ApJ...542..914W} for details). We can then consider
our estimates (obtained by assuming a solar composition) as
conservative. A detailed description of the local absorption,
however, would requires a detailed knowledge of the distribution
of the chemical abundances in the ejecta, which is beyond the scope
of this paper.

\section{Summary and concluding remarks}
\label{sec4}

We investigated how the morphology and the emission properties of the
remnant of SN\,1987A reflect: 1) the physical characteristics of the
progenitor SN and 2) the early interaction of the SN blast wave with the
surrounding inhomogeneous nebula. To this end, we have developed a model
describing SN\,1987A from the breakout of the shock wave at the stellar
surface up to the 3D expansion of its remnant. A major challange was
capturing the enormous range in spatial scales (spanning six orders of
magnitude) that required a very high spatial resolution (down to $\sim
0.2$~AU). We performed an exploration of the parameters space, searching
for a model reproducing at the same time both the observables of the SN
(i.e. bolometric lightcurve, evolution of line velocities, and continuum
temperature at the photosphere) and the X-ray emission of the remnant
(lightcurves, spectra, and morphology). Our findings lead to several
conclusions:

1. We identified three phases in the evolution (see Fig.\ref{fig2} and
on-line movie). During the first phase (H\,II-region-dominated
phase; from year 3 to 15) the fastest ejecta interact with the H\,II
region and the X-ray emission is dominated by shocked plasma from
this region and by a smaller contribution from the outermost ejecta
(see Figs.~\ref{fig1}, \ref{fig3}, and on-line movie). In
the second phase (ring-dominated phase; from year 15 to 32) the ejecta
interacts with the dense equatorial ring.  The emission in the soft X-ray
band is largely dominated by the shocked clumps (see Figs.~\ref{fig1},
\ref{fig3}, and on-line movie). The emission in the hard X-ray
band is mainly due to shocked plasma from the smooth component of the
ring and to fragments of the shocked clumps stripped by hydrodynamic
instabilities developing at the cloud boundaries.  In the third phase
(ejecta-dominated phase; from year 32), the forward shock propagates
beyond the majority of the dense ring material and the reverse shock
travels through the inner envelope of the SN. The X-ray emission is
dominated by shocked ejecta (see Figs.~\ref{fig1} and \ref{fig3}) and
the remnant morphology is characterized by very bright knots due to
shocked clumps of ejecta (see on-line movie).

2. Our favored model reproduces altogether the main observables of
SN\,1987A (bolometric lightcurve and evolution of photospheric temperature
and velocity) during the first 250 days after explosion and the observed
X-ray lightcurves and spectra of its remnant (see Figs.~\ref{fig1}
and \ref{fig3}) in the following 30 years. Therefore, we have
demonstrated that the physical model reproducing the observables of the
SN is able to reproduce also the observables of the subsequent expanding
remnant, providing a firm link between two research fields (SN explosions
and SNR evolution) which, traditionally, are based on models that are
independent from each other. In other words, in the case of SN\,1987A,
we have demonstrated the consistency between the cause (the SN explosion)
and the effect (the interaction of the remnant with the surrounding
medium). This is a great advance with respect to a parameterised explosion
model to bootstrap the SNR (as those commonly used in the literature)
whose parameters are chosen to fit the observables of the remnant but
that does not ensure to fit also the observables of the progenitor SN.

3. From our favored model, we identified the imprint of SN\,1987A on
the remnant emission.  In particular, we constrained the SN explosion
energy in the range $1.2-1.4\times 10^{51}$~erg and the envelope mass
in the range $15-17 M_{\odot}$. The model also constrained the physical
properties of post-explosion ejecta. During the H\,II-region-dominated
phase, the sudden increase of X-ray flux observed around year 4
(Fig.~\ref{fig1}b) is reproduced if the outermost ejecta have a
post-explosion radial profile of density approximated by a power law
with index $\alpha = -8$. On the contrary, models with index $\alpha =
-9$ (as early suggested for SN\,1987A; \citealt{1988ApJ...330..218W})
systematically underestimate the soft X-ray flux during the first 7
years since the explosion (see Fig.~\ref{prof_n9}), independently of the
density structure of the nebula within the range of values compatible
with observations.  Indeed the shape of the lightcurves in this phase
reflects the structure of outer ejecta and reveals the imprint of the
SN on the remnant emission.

4. Our favored model allowed us to constrain the structure of the pre-SN
nebula. In the ring-dominated phase, the shape of lightcurves and spectra
reflect the density structure of the nebula, allowing to disentangle
the effects of the SN event (identified in the previous phase) from
those of remnant interaction with the environment. This enabled us to
ascertain the origin of the multi-thermal X-ray emission and to constrain
the nebula structure (see Table~\ref{tab1} for all the details). From
this model we estimated that the total mass of the ring is $M_{\rm rg} =
0.062\,M_{\odot}$ of which $\sim 64$\% is plasma with density $n \approx
1000$~cm$^{-3}$ and $\sim 36$\% is plasma with $n \approx 2.5\times
10^4$~cm$^{-3}$.  These values are in excellent agreement with those
derived from optical spectroscopic data for the density structure of
the ionized gas of the ring (\citealt{2010ApJ...717.1140M}).

Our model enabled us to make predictions about the ejecta evolution
and the future changes in the remnant morphology in X-rays. In the
next few years, the remnant will enter in the ejecta-dominated
phase. The X-ray flux will reflect the radial profile of density
in outer ejecta: the steeper the slope of this profile, the higher
the emission from shocked ejecta. The emission will enable us to
study in more details the ejecta asymmetries and the distribution
of metal-rich layers. This will provide important clues on the
dynamics of the explosion and might even improve our knowledge about
the nucleosynthesis processes occurring in the latest stage of
stellar evolution and during the SN explosion, making the remnant
of SN\,1987A a unique probe of CC-SNe.

Finally, we investigated how our model relate to the
existence of the yet undetected central compact remnant. The complete
picture of the line of sight column towards the center of SN\,1987A
provied by our model has shown that the emission of the compact
remnant cannot be revealed yet due to a local absorbing column
density which is a factor of 20 higher than the interstellar
absorption toward the Large Magellanic Cloud. The constraint on the
thermal emission from the central source inferred from our model
is in good agreement with standard theories of neutron star cooling.

\acknowledgments
We thank Roger Chevalier, Dan Dewey, Fabio Reale and Salvo Sciortino
for helpfull discussions. We also thank the anonymous referee for
useful suggestions that we have incorporated into the paper. This
paper was partially funded by the PRIN INAF 2014 grant. MLP
acknowledges financial support from CSFNSM and from INAF. The
software used in this work was, in part, developed by the U.S.
Department of Energy–supported Advanced Simulation and Computing/Alliance
Center for Astrophysical Thermonuclear Flashes at the University
of Chicago. We acknowledge that the results of this research have
been achieved using the PRACE Research Infrastructure resource
MareNostrum III based in Spain at the Barcelona Supercomputing
Center (PRACE Award N.2012060993). We acknowledge the CINECA Award
N.HP10BI36DG,2012 for the availability of high performance computing
resources and support.

\bibliographystyle{apj}
\bibliography{biblio}

\newpage
\appendix

\section{Effect of radioactive heating during the remnant expansion}
\label{SNSNRcomparison}

\begin{figure}[!t]
  \centering
  \includegraphics[width=8.5cm]{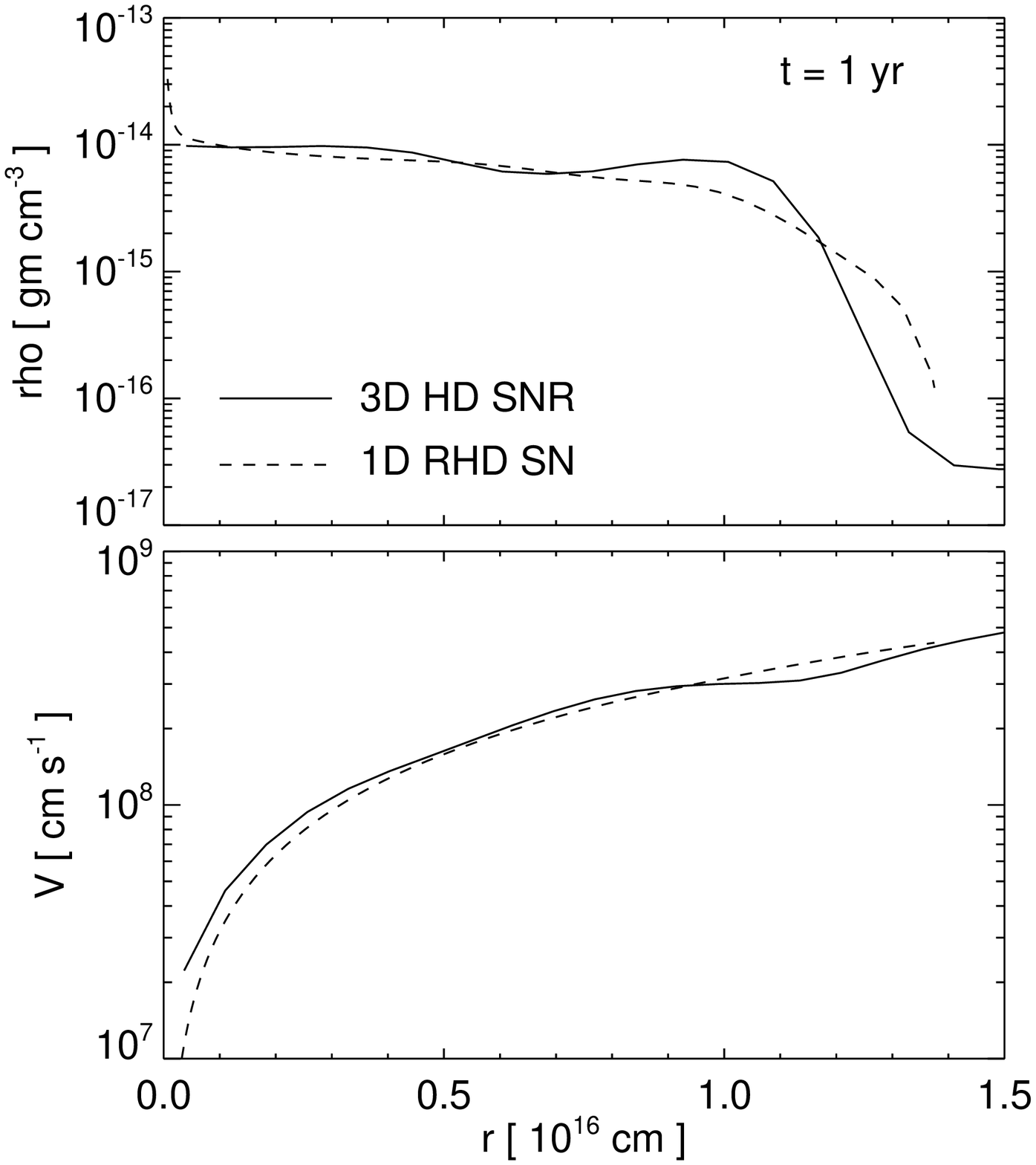}
  \caption{Radial profiles of mass density (upper panel) and velocity
  (lower panel) of ejecta at t = 1 yr since the SN event. The solid
  curves show the angle-averaged profiles derived with the 3D
  hydrodynamic model describing the SNR evolution; the dashed curves
  show the analogous profiles derived with the 1D relativistic
  radiation hydrodynamic code describing the post-explosion evolution
  of the SN.}
\label{figapp}
\end{figure}

The 3D hydrodynamic simulations describing the evolution
of the SNR and its interaction with the inhomogeneous CSM do not
include a heating term due to decays of radioactive isotopes
synthesized in the SN explosion (e.g. $^{56}$Co or $^{44}$Ti). To
check the validity of our assumption, we used our 1D relativistic
radiation hydrodynamic code (which includes the radioactive heating
term; see Sect.~\ref{modelSN}), to extend run SN-M17-E1.2-N8,
simulating the post-explosion evolution of the SN, and covering the
first year of evolution. Then we compared the radial profiles of
mass density and velocity of ejecta derived in this case at $t=
1$~yr with the angle-averaged radial profiles of density and velocity
derived at the same epoch with our 3D hydrodynamic model (not
including the radioactive heating). Fig.~\ref{figapp} shows some
differences in the density profiles that are mainly due to the
effect of ejecta clumping considered only in the 3D simulation. On
the other hand, the two velocity profiles are very similar suggesting
that the effect of radioactive heating does not affect too much the
dynamics of the ejecta. This result suggests that we can safely
neglect the effect of radioactive heating during the interaction
of the remnant with the surrounding nebula.

\section{The data analysis}
\label{datanalysis}

We present new results of the analysis of different observations
of the remnant of SN\,1987A performed with the {\it ASCA},
{\it XMM-Newton}, and {\it Chandra} X-ray telescopes. We adopted Xselect
V2.4, CIAO V4.6.1, and SAS V13 for the reduction of {\it ASCA}/SIS,
{\it XMM-Newton}/EPIC, and {\it Chandra}/ACIS data, respectively. All
the spectra were analyzed with XSPEC V12.8.2.

\subsection{{\it ASCA} data}

We analyzed the {\it ASCA} observations ID 55039000 (performed on
November 1997), 56041000 (November 1998), and 57034000 (November
1999), to derive the X-ray unabsorbed fluxes in the $[0.5,2]$ keV
and $[3,10]$ keV bands, shown in Fig.~\ref{fig1}. We screened the data
according to the standard screening criteria and we added the
screened SIS0 and SIS1 spectra by using the ADDASCASPEC task. We
verified that the spectra extracted from observations 55039000 and
56041000 were consistent with each other, considering the
relatively low signal-to-noise ratio of the early observations of
SN\,1987A. We then fitted the SIS spectra extracted from observations
55039000 and 56041000 simultaneously. For each spectrum, we subtracted
a background spectrum extracted from a nearby region immediately
outside of the supernova remnant (SNR) shell and we verified that
the best-fit values do not depend significantly on the choice of
the background region.  Spectra were modeled by an absorbed (TBABS
model in XSPEC) optically thin plasma in non-equilibrium of ionization
(VNEI model). In the fittings, the plasma chemical abundances were
fixed to the values derived from the analysis of deep {\it Chandra}/LETG
and HETG observations of SN\,1987A (\citealt{2009ApJ...692.1190Z}), while
the interstellar column density was fixed to (\citealt{2006ApJ...646.1001P})
$N_{\rm H} = 2.35\times 10^{21}$ cm$^{-2}$. Our best-fit models
provide a good description of the spectra ($\chi^2=41.5$ with 47
d.o.f. for the 1997 and 1998 spectra, and $\chi^2=36.0$ with 29
d.o.f. for the 1999 spectrum) and the best-fit temperatures and
ionization timescales are $kT=1.5\pm0.3$ keV,
$\tau=3\pm2\times10^{10}$ s cm$^{-3}$ for the 1997 and 1998
spectra and $kT=1.7\pm0.4$ keV, $\tau=2\pm1\times10^{10}$ s
cm$^{-3}$ for the 1999 spectrum.

\subsection{{\it XMM-Newton} data}

As for the {\it XMM-Newton} data, we analyzed observation ID 0083250101
(performed on April 2001) and the previously unpublished observation
ID 0690510101 (performed on December 2012). We focussed on the EPIC
data and we selected events with PATTERN$\le12$ for the MOS cameras,
PATTERN$\le$4 for the pn camera, and FLAG=0 for both. We screened the
original event files by using the sigma-clipping algorithm (ESPFILT
tasks). The pn spectra extracted from the 2001$/$2012 observations are
shown in the upper$/$middle panels of Fig.~\ref{fig3}. To derive the $[0.5,2]$
keV and $[3,10]$ keV fluxes of the 2012 observation, we fitted the pn
and MOS spectra simultaneously (we selected a circular extraction region
with radius $r=20''$) in the $[0.3,9]$ keV energy band. We adopted a model
including three components of optically thin plasma in non-equilibrium of
ionization (NEI), widely used (\citealt{2012A&A...548L...3M}) to describe the
latest X-ray observations of SN\,1987A.  Again we fixed the interstellar
column density (\citealt{2006ApJ...646.1001P}) to $N_{\rm H} = 2.35\times
10^{21}$ cm$^{-2}$.  The chemical abundances were fixed to the values
derived in previous studies  (\citealt{2009ApJ...692.1190Z}) for all the three
components. The best-fit temperatures are $kT_1=0.48^{+0.01}_{-0.02}$
keV, $kT_2=0.77\pm0.03$ keV, and $kT_3=2.41\pm0.05$ keV, while the
corresponding ionization timescales are $\tau_1=1.5\pm0.1\times10^{11}$
s cm$^{-3}$, $\tau_2=2.4\pm0.3\times10^{13}$ s cm$^{-3}$, and
$\tau_3=1.11^{+0.05}_{-0.03} \times10^{11}$ s cm$^{-3}$. The presence of
a component with a $\tau>10^{13}$  s cm$^{-3}$ (in our case, component
2) indicates that part of the X-ray emitting plasma has reached the
collisional ionization equilibrium and is in agreement with previous
findings (\citealt{2013ApJ...764...11H}). We point out that these
models only provide heuristic (phenomenological) descriptions of
the spectra, whose physical origin can be unveiled only by accurate
hydrodynamic modeling (see Section~\ref{evolSNR}).

\subsection{{\it Chandra} data}

Finally, we analyzed the {\it Chandra}/ACIS observation 1967
(performed on December 2000) and the previously unpublished observation
14697 (March 2013) to produce the X-ray images shown in Fig.~\ref{fig2}
(lower panels). To study the morphology of SN\,1987A from the
observations, we carefully followed the procedure described in the
literature (\citealt{2000ApJ...543L.149B, 2002ApJ...567..314P}). The
ACIS data were deconvolved with a maximum likelihood algorithm (with
25 iterations; \citealt{1972JOSA...62...55R,1974AJ.....79..745L}), using an
on-axis point-spread function produced by the MARX simulation package. The
high photon statistics allowed us to use $0.062''$ pixels for the
deconvolution process. The deconvolved images were finally smoothed
with a Gaussian with $\sigma=0.1''$.

\end{document}